\documentclass[12pt, preprint]{aastex}
\usepackage{natbib}
\usepackage{graphicx}
\usepackage{amsmath}
\usepackage{amssymb}
\usepackage{mathrsfs}
\usepackage[utf8]{inputenc}
\usepackage{longtable}

\shorttitle{CRRLs II}
\shortauthors{Salgado et al.}

\begin{document}
\title{Low Frequency Carbon Radio Recombination Lines II: The Diffuse Interstellar Medium}

\author{F.~Salgado\altaffilmark{1}, L.~K.~Morabito\altaffilmark{1}, J.~B.~R.~Oonk\altaffilmark{1,2}, P. Salas\altaffilmark{1},
M.~C.~Toribio\altaffilmark{1}, H.~J.~A.~R\"ottgering\altaffilmark{1}, A.~G.~G.~M.~Tielens\altaffilmark{1}}

\altaffiltext{1}{Leiden Observatory, University of Leiden, P. O. Box 9513, 2300 RA Leiden, Netherlands}
\altaffiltext{2}{Netherlands Institute for Radio Astronomy (ASTRON), Postbus 2, 7990 AA Dwingeloo, The Netherlands}

\begin{abstract}
In the second paper of the series, we have modeled low frequency carbon radio recombination lines (CRRL) from the interstellar medium.
Anticipating the LOw Frequency ARray (LOFAR) survey of Galactic CRRLs, we focus our study on the physical conditions of the diffuse cold neutral medium (CNM).
We have used the improved departure coefficients computed in the first paper of the series to calculate line-to-continuum ratios. The results show that the
line width and integrated optical depths of CRRL are sensitive probes of the electron density, gas temperature, and the emission measure of the cloud. Furthermore,
the ratio of CRRL to the [CII] at 158~$\mu$m line is a strong function of the temperature and density of diffuse clouds. Guided by our calculations, we
analyze CRRL observations and illustrate their use with data from the literature. 
\end{abstract}

\maketitle

\section{Introduction}

The interstellar medium (ISM) plays a central role in the evolution of galaxies. The formation of new stars slowly consumes the ISM,
locking it up for millions to billions of years while stars, as they age, return much of their mass increased in metallicity, back to the ISM.
Stars also inject radiative and kinetic energy into the ISM and this controls the physical characteristics (density, temperature and pressure)
as well as the dynamics of the gas as revealed in observed spectra. This interplay of stars and surrounding gas leads to the presence of
distinct phases (e.g. \citealt{field1969,mckee1977}). Diffuse atomic clouds (the Cold Neutral Medium, CNM) have densities of about $50~\mathrm{cm^{-3}}$ and temperatures
of about $80~\mathrm{K}$, where atomic hydrogen is largely neutral but carbon is singly ionized by photons with energies between $11.2~\mathrm{eV}$ and $13.6~\mathrm{eV}$.
The warmer ($\sim8000~\mathrm{K}$) and more tenuous ($\sim0.5~\mathrm{cm^{-3}}$) intercloud phase [the Warm Neutral medium (WNM)
and Warm Ionized Medium (WIM)] is heated and ionized by FUV and EUV photons escaping from HII regions \citep{wolfire2003}.
While these phases are often considered to be in thermal equilibrium and in pressure balance, the observed large turbulent width and
presence of gas at thermally unstable, intermediate temperatures may indicate that kinetic energy input is important. Thermally unstable gas
could indicate that the gas does not have sufficient time to cool between subsequent passages of a shock or after intermittent dissipation of turbulence
(e.g. \citealt{kim2011}). In addition, the ISM also hosts molecular clouds, where hydrogen is in the form of $\mathrm{H_2}$
and self-gravity plays an important role. All of these phases are directly tied to key questions on the origin and evolution of the ISM,
including energetics of the CNM, WNM and the WIM; the evolutionary relationship of atomic and molecular gas; the relationship of these ISM phases
with newly formed stars; and the conversion of their radiative and kinetic power into thermal and turbulent energy of the ISM (e.g. \citealt{cox2005,elmegreen2004, scalo2004,mckee2007}).

The diffuse interstellar medium has been long studied using, in particular, the 21 cm hyperfine transition of neutral atomic hydrogen (e.g. \citealt{kulkarni1987,heilesandtroland2003a}).
These observations have revealed the prevalence of a two phase structure in the interstellar medium of cold clouds embedded in a warm intercloud medium. However, it has been notoriously difficult to determine
the physical characteristics (density, temperature) of these structures in the ISM as HI by itself does not provide a good probe. Optical and UV observations of atomic lines can provide
the physical conditions but are by necessity limited to pinpoint experiments towards bright background sources. However, with the opening up of the low frequency
radio sky with modern interferometers such as the Low Frequency ARray for Radioastronomy (LOFAR, \citealt{vhaarlem2013}), Murchison Wide field Array \citep{tingay2013},
Long Wavelength Array \citep{ellingson2013} and, in the future, the Square Kilometer Array (SKA), systematic surveys of low frequency ($\nu \lesssim 300~\mathrm{MHz}$)
Carbon Radio Recombination Lines (CRRLs) have come in reach and these surveys can be expected to quantitatively measure the conditions in the emitting
gas \citep{oonk2015a}.

Carbon has a lower ionization potential (11.2 eV) than hydrogen and can be ionized by radiation fields in regions where hydrogen is largely neutral. Recombination of carbon ions
with electrons to high Rydberg states will lead to CRRLs in the sub-millimeter to decameter range. CRRLs have been observed in the interstellar medium of
our Galaxy towards two types of clouds: diffuse clouds (e.g.: \citealt{konovalenko1981, erickson1995, roshi2002, stepkin2007,oonk2014}) and
photodissociation regions (PDRs), the boundaries of HII regions and their parent molecular clouds (e.g.: \citealt{natta1994, wyrowski1997, quireza2006}).
Recently, \citet{morabito2014} discovered extragalactic CRRLs associated with the nucleus of the nearby starburst galaxy, M82. Theoretical models for CRRLs
were first developed by \citet{watson1980} and \citet{walmsley1982}, including the effects of dielectronic recombination \footnote{As in \citet{salgado2015}, following common usage in the astronomical literature,
we refer to this process as dielectronic recombination rather than the more appropriate dielectronic capture.} with the simultaneous excitation of the
${^2}P_{3/2}$ fine-structure level and later extended by \citet{ponomarev1992} and by \citet{payne1994}. However, these studies were hampered by the limited computer
resources available at that time. 

In the coming years, we will use LOFAR to carry out a full northern hemisphere survey of CRRL emitting clouds in the Milky Way. This will allow us to study the thermal balance,
chemical enrichment and ionization rate of the cold neutral medium from degree-scales down to scales corresponding to individual clouds and filaments in our Galaxy. Furthermore,
following the first detection of low-frequency CRRLs in an extragalactic source (M82; \citealt{morabito2014}) we will also use LOFAR to perform the first flux limited survey
of CRRLs in extragalactic sources. Given the renewed observational interest in CRRLs, a new theoretical effort seems warranted. In the first paper of this series,
(\citealt{salgado2015}, hereafter Paper~I), we studied the level population of hydrogenic atoms including the effects of dielectronic recombination in carbon atoms. The level population of atoms,
however, is not the only process that influences the strength of an observed line as radiative transfer effects can alter the strength/depth of an observed line. In this paper,
we use the results of Paper~I to develop CRRLs as a tool to derive the physical conditions in the emitting gas. In this, we will focus on cold diffuse clouds as these are
expected to dominate the low frequency CRRL sky. The paper is organized as follows: in Section \ref{section_radtransf} we review radiative transfer theory in the context of radio
recombination lines. We review the line broadening mechanisms of CRRLs in Section \ref{section_lineprofile}. In Section \ref{section_results}, we present the results of our
models and compare them with observations from the literature and provide guidelines to analyze such observations.
Finally, in Section \ref{section_conclusions}, we summarize our results and provide the conclusions of our work.

\section{Theory}

\subsection{Radiative transfer of carbon radio recombination lines}\label{section_radtransf}

The physical conditions of the diffuse interstellar medium (temperatures of $T_e\approx 100~\mathrm{K}$ and electron densities $N_e\approx 10^{-2}~\mathrm{cm^{-3}}$) favor an increase in
the level population at high quantum levels via dielectronic recombination (Paper~I). Moreover, the presence of an external radiation field can also alter the level population
of carbon atoms. In addition, while low frequency CRRLs are observed in absorption against a background continuum
(e.g. \citealt{kantharia2001,oonk2014,morabito2014}), high frequency recombination lines are observed in emission. Therefore, radiative transfer effects must be analyzed in
order to derive meaningful physical parameters from observations.

We begin our analysis by revisiting the radiative transfer problem in the context of CRRLs. At a given frequency, the observed emission has
two components, corresponding to the line transition itself and the underlying continuum emission. In Appendix \ref{appendix_radtransfer},
we summarize the standard general solution to the one dimensional radiative transfer equation of a line in a homogeneous medium. Here, we show the result
for a cloud at a constant temperature $T_e$ \footnote{Throughout this article we assume a filling factor of unity.}:
\begin{eqnarray}\label{eq_linecontgeneral}
\frac{I_\nu^{line}}{I_\nu^{cont}} &=& \frac{\eta B_\nu(T_e) (1-e^{-\tau_\nu^{total}})+I_0(\nu) e^{-\tau_\nu^{total}}}{B_\nu(T_e)(1-e^{-\tau_\nu^c}) + I_0(\nu) e^{-\tau_\nu^c}} -1,
\end{eqnarray}
\noindent where $I_\nu^{line}$ is the intensity of a line at a frequency $\nu$, $I_\nu^{cont}$ is the intensity of the continuum, $\eta$ is a
correction factor to the Planck function due to non-LTE effects (as defined in \citealt{strelnitski1996,gordon2009}, see Appendix \ref{appendix_radtransfer}), $B_\nu(T_e)$
is the Planck function, $\tau_\nu^{total}$ is the sum of the line and continuum optical depth (${\tau_\nu^l~\mathrm{and}~\tau_\nu^c}$, respectively) and
$I_0(\nu)$ is the intensity of a background continuum source at the frequency of the line \footnote{In Appendix A, we provide a comprehensive list of the symbols used in this article.}.

In the presence of a strong background radiation field, as is the case for low frequency lines in the diffuse ISM ($I_0 \gg \eta B_\nu(T_e)$, see below),
the background term ($I_0$) dominates and the first term in the numerator and denominator on the right-hand-side of Equation~\ref{eq_linecontgeneral}
can be ignored and this equation simplifies to,
\begin{eqnarray}\label{eq_linecontapp}
\frac{I_\nu^{line}}{I_\nu^{cont}}&=& e^{-\tau_\nu^l}-1,
\end{eqnarray}
\noindent independent of the background source. Assuming that the line is optically thin ($|\tau_\nu^l| \ll1$), Equation~\ref{eq_linecontapp} is
approximated by (e.g. \citealt{kantharia2001}):
\begin{eqnarray}\label{eqn_lintoconttau}
\frac{I_\nu^{line}}{I_\nu^{cont}}&=& -\tau_\nu^l.
\end{eqnarray}
\noindent Note that, due to the minus sign on the right hand side of Equation~\ref{eqn_lintoconttau},
when $\tau_l$ is positive the line is observed in absorption against the background source.

From the definition of $\tau_\nu^l$ (see Appendix \ref{appendix_radtransfer}) and explicitly considering the normalized line profile, $\phi(\nu),~(\mathrm{with}~\int{\phi(\nu)}=1)$:
\begin{eqnarray}\label{eqn_linetocontapp}
\frac{I_\nu^{line}}{I_\nu^{cont}}&=& - \kappa_\nu^l \phi(\nu) L.
\end{eqnarray}
Introducing the departure coefficients from LTE, $b_n$ and the correction factor for stimulated emission or absorption, $\beta_n$ \citep{brocklehurst1972,gordon2009}, we can write,
\begin{eqnarray}
\frac{I_\nu^{line}}{I_\nu^{cont}}&=& - \kappa_\nu^l(LTE) \phi(\nu)  b_n \beta_{nn'}  L,\nonumber\\
\frac{I_\nu^{line}}{I_\nu^{cont}}&\approx& -1.069\times10^7 \Delta n M(\Delta n)  \frac{b_n \beta_{nn'}}{T_e^{5/2}} e^{\chi_n} EM_{\mathrm{C+}} \phi(\nu),
\end{eqnarray}
\noindent assuming $h \nu \ll kT_e$, and $\Delta n/n \ll 1$, in Equation 5 we have inserted the value for $\kappa_\nu^l$ absorption coefficient (Appendix \ref{appendix_radtransfer}).
Here, $EM_{\mathrm{C+}}=N_e N_{\mathrm{C+}} L$ is the emission measure in units of $\mathrm{cm^{-6}~pc}$, $N_e$ is the electron density, $N_{C+}$ is the carbon ion density and $L$ is the
pathlength of the cloud in pc. $\Delta n=n'-n$ is the difference between the levels involved in the transition,
the factor $M(\Delta n)$\footnote{Some values for $M(\Delta n)=0.1908,~0.02633,~0.008106,~0.003492,~0.001812$, for $\Delta n =1,~2,~3,~4,~5$, respectively \citep{menzel1968}.}
comes from the approximation to the oscillator strength of the transition, as given by \citet{menzel1968} (see Appendix \ref{appendix_radbroadening}). The $b_n\beta_{nn'}$ factor relates
the line emission or absorption to the level population of the emitting atoms and has been calculated following the method described in Paper~I;
${\chi_n= hc Z^2 Ry /n^2 k T_e}$, as defined in Appendix B.

At the line center, the line to continuum ratio depends on the broadening of the line (see Section \ref{section_lineprofile}, below). However, we can remove
the dependence on the line profile by integrating the line over frequency:
\begin{eqnarray}\label{eqn_integlinetocont}
\int \frac{I_\nu^{line}}{I_\nu^{cont}} \mathrm{d}\nu&=& -1.069\times10^7 \Delta n M(\Delta n) \frac{b_n \beta_{nn'}}{T_e^{5/2}} e^{\chi_n} EM_{\mathrm{C+}}~\mathrm{Hz}.
\end{eqnarray}
Note that by setting $\Delta n=1$ (i.e. for $\mathrm{C}n\alpha$ lines\footnote{We will refer to electron transitions in carbon from levels $n+1\rightarrow n$ as $\mathrm{C}n\alpha$, $n+2\rightarrow n$ as $\mathrm{C}n\beta$
and $n+3\rightarrow n$ as $\mathrm{C}n\gamma$ \citep{gordon2009}.}) in Equation~\ref{eqn_integlinetocont} we recover Equation~70 in \citet{shaver1975}
and Equation~5 in \citet{payne1994}.

For high densities, $b_n\beta_n$ approaches unity at high $n$ levels and the integrated line to continuum ratio changes little with $n$ for a
given $T_e$ and $EM_{\mathrm{C+}}$. When the $\beta_n$ factor in Equation~\ref{eqn_integlinetocont} is positive (negative) the line is in absorption (emission).
The strong dependence on electron temperature of the integrated line to continuum ratio ($\propto T_e^{-2.5}$) favors the detection of low temperature clouds.
An increase of a factor of two (three) in the temperature reduces the integrated line to continuum by a factor of about 6 (15), all other terms being equal.

From Equation~\ref{eqn_integlinetocont}, we note that for $\mathrm{C}n\alpha$ lines: 
\begin{eqnarray}\label{eqn_integlinetocontx}
\int \frac{I_\nu^{line}}{I_\nu^{cont}} \mathrm{d}\nu&=& -20.4 b_n \beta_{nn'}\left(\frac{T_e}{100~\mathrm{K}}\right)^{-2.5} EM_{\mathrm{C+}} ~\mathrm{Hz},\\
 &=&-0.2 b_n \beta_{nn'}\left(\frac{T_e}{100~\mathrm{K}}\right)^{-2.5} \left(\frac{N_e}{0.1~\mathrm{cm^{-3}}}\right)^2  \left(\frac{L}{\mathrm{pc}}\right)~\mathrm{Hz},
\end{eqnarray}
\noindent assuming that electrons are produced by singly ionized carbon ($N_e=N_{C+}$) and for high $n$ level ($n\gg \sqrt(1.6\times10^5/T_e)$). The typical optical depths that can be observed
with current instruments are $\sim 10^{-3}$. As we already mentioned, for high $n$ $b_n\beta_n\simeq 1$. Hence, clouds ($L\simeq 5$~pc) with electron densities
greater than $10^{-2}$ cm$^{-3}$ (hydrogen densities $>50$ cm$^{-3}$) are readily observable.

\subsection{The far-infrared fine structure line of $\mathrm{C+}$}

The fine structure transition ${^2}P_{1/2}-{^2}P_{3/2}$ of carbon ions is one of the main coolants in diffuse neutral clouds at 158~$\mu\mathrm{m}$.
Moreover, the [CII] 158 $\mu\mathrm{m}$ is directly linked to the level population of carbon atoms at low temperatures
through the dielectronic recombination process. In Section \ref{section_results}, we show how observations of this line combined with CRRLs can
be used as powerful probes of the temperature of diffuse neutral clouds. Here, we give a description of an emission model of the line.
The intensity of the [CII] 158 $\mu\mathrm{m}$~line in the optically thin limit is given by (e.g. \citealt{sorochenko2000}):
\begin{eqnarray}
I_{158}&=& \frac{h\nu}{4 \pi} A_{3/2,1/2} N^+_{3/2} L \nonumber\\
       &=& \frac{h\nu}{4 \pi} \frac{A_{3/2,1/2} 2\exp(-92/T_e) R}{1+2 \exp(-92/T_e) R}N_{\mathrm{C^+}} L,
\end{eqnarray}
\noindent with $\nu$ the frequency of the ${^2}P_{1/2}-{^2}P_{3/2}$ transition, $A_{3/2,1/2}=2.4\times10^{-6}~\mathrm{s^{-1}}$
is the spontaneous transition rate, $N^+_{3/2}$ the number density of carbon ions in the $3/2$ state, $L$ the length along
the line of sight of the observed cloud and $N_{\mathrm{C^+}}$ the density of carbon ions; $R$ is defined in \citet{ponomarev1992,payne1994} (see Paper~I):
\begin{eqnarray}\label{eqpayner}
R = \frac{N_e\gamma_e+N_H\gamma_H}{N_e\gamma_e+N_H\gamma_H+A_{3/2,1/2}},
\end{eqnarray}
\noindent where $\gamma_e~\mathrm{and}~\gamma_H$ are the de-excitation rates due to electrons and hydrogen atoms, respectively. The rates involved 
are detailed in Paper~I. We assume that collisions with electrons and hydrogen atoms dominate over molecular hydrogen and neglect collisions with H$_2$, as in Paper~I.
This is a good approximation for diffuse clouds with column densities up to $\sim10^{21}$ cm$^{-2}$. For larger column densities, the H/H$_2$ transition will have to be modeled
in order to evaluate $R$.

The optical depth of the C$+$ fine structure line for the transition ${^2}P_{1/2}-{^2}P_{3/2}$ is given by \citet{crawford1985,sorochenko2000}:
\begin{eqnarray}
\tau_{158} &=& \frac{c^2}{ 8 \pi \nu^2} \frac{A_{3/2,1/2}}{1.06 \Delta \nu} 2 \alpha_{1/2} \beta_{158} N_{\mathrm{C^+}} L,
\end{eqnarray}
\noindent $\Delta \nu$ is the FWHM of the line (assumed to be Gaussian); the $\alpha_{1/2}(T_e)$ and $\beta_{158}(T_e)$ coefficients depend on the electron
temperature of the cloud and are defined by \citet{sorochenko2000}:
\begin{eqnarray}
\alpha_{1/2}(T_e) &=& \frac{1}{1+2 \exp(-92/T_e) R},\\
\beta_{158}(T_e) &=& 1-\exp(-92/T_e)R.
\end{eqnarray}
Adopting a line width of $2~\mathrm{km~s^{-1}}$, at low electron temperatures and densities, the FIR [CII] line is optically thin for hydrogen column densities less than
about $1.2\times10^{21}~\mathrm{cm^{-2}}$. For a cloud size of 5~pc, this corresponds to hydrogen densities of $\sim 10^2$~cm$^{-3}$ and electron densities of $\simeq 10^{-2}$~cm$^{-3}$
if carbon is the dominant ion.

\subsection{Line profile of recombination lines}\label{section_lineprofile}

The observed profile of a line depends on the physical conditions of the cloud, as an increase in electron density and temperature or the presence of a radiation field
can broaden the line and this is particularly important for high $n$.
Therefore, in order to determine the detectability of a line, the profile must be considered. Conversely, the observed line width of recombination lines provides additional
information on the physical properties of the cloud.

The line profile is given by the convolution of a Gaussian and a Lorentzian profile, and is known as a Voigt profile \citep{shaver1975,gordon2009}.
Consider a cloud of gas of carbon ions at a temperature $T_e$. Random thermal motions of the atoms in the gas produce shifts in frequency that
reflect on the line profile as a Gaussian broadening (Doppler broadening). In the most general case, turbulence can increase the width of a line and, as is common in the
literature (e.g. \citealt{rybicki1986}), we describe the turbulence by an RMS turbulent velocity. Thus, the Gaussian line profile can be described by:
\begin{eqnarray}
\Delta \nu_D= \frac{\nu_0}{c}\sqrt{\frac{2kT_e}{m_C}+\langle v_{RMS} \rangle^2},
\end{eqnarray}
\noindent where $m_C$ is the mass of the carbon atom and $\langle v_{RMS} \rangle$ is the RMS turbulent velocity. The Gaussian width in frequency space
is proportional to the frequency of the line transition. 

At low frequencies, collisions and radiation broadening dominate the line width.
The Lorentzian (FWHM) broadening produced by collisions is given by:
\begin{eqnarray}
\Delta \nu_{col} = \frac{1}{\pi}\sum_{n\neq n'} N_e C_{n'n},
\end{eqnarray}
\noindent where $C_{n'n}$ is the collision rate for electron induced transitions from level $n'$ to $n$, and
$N_e$ is the electron density. Note that $C_{n'n}$ depends on temperature \citep{shaver1975,gordon2009}. In order to estimate the collisional broadening,
we fitted the following function at temperatures between 10 and 30000~K:
\begin{eqnarray}\label{eqn_fitcol}\label{eqn_colbroaden}
\sum_{n\neq n'} C_{n'n} =   10^{a(T_e)} n^{\gamma_c(T_e)},
\end{eqnarray}
\noindent which is valid for levels $n > 100$. Values for $a(T_e)$ and $\gamma_c(T_e)$ as a function of electron temperature are given in Table \ref{table_col}.

\begin{figure}[!ht]
\includegraphics[width=1\columnwidth]{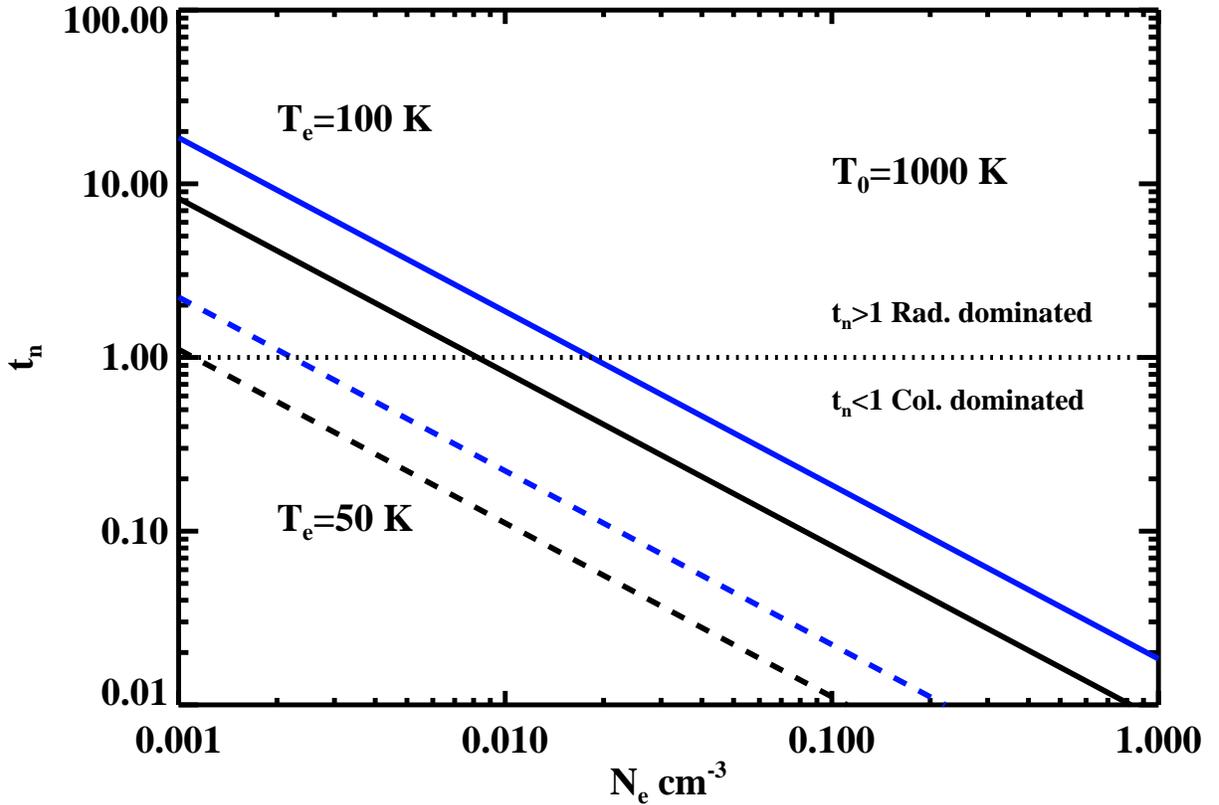}
\caption{The $t_n$ factor defined in Equation~\ref{eqn_tnfac} as a function of electron density for quantum $n$ levels between 200 (black line) and 1000 (blue line).
The figure is presented for two electron temperatures: $T_e=50~\mathrm{K}$ (dashed lines) and $T_e=100~\mathrm{K}$ (solid lines). The dotted
line marks the boundary for the line widths being in the collision dominated regime ($t_n<1$) and the radiative dominated regime ($t_n>1$).\label{fig_tnfac}}
\end{figure}

In a similar way as for collisional broadening, the interaction of an emitter with a radiation field produces a broadening of the line profile.
In Appendix \ref{appendix_radbroadening}, we give a detailed expansion for different external radiation fields. Here, we discuss the case of a
synchrotron radiation field characterized by a power-law with a temperature $T_0$ at a reference frequency $\nu_0=100~\mathrm{MHz}$ and an spectral
index $\alpha_{pl}=-2.6$ (see section~\ref{section_results}). Under the above considerations, the FWHM for radiation broadening is given by :
\begin{eqnarray}\label{eqn_radbroaden}
\Delta \nu_{rad}= 6.096\times10^{-17} T_0 n^{5.8}~(\mathrm{s^{-1}}).
\end{eqnarray}
As is the case for collisional broadening, radiation broadening depends only on the level and the strength of the surrounding radiation field. The dependence
on $n$ is stronger than that of collisional broadening at low densities and radiation broadening dominates over collisional broadening. As the density decreases,
the level $n$ where radiation broadening dominates decreases. In order to estimate where this occurs we define $t_n$ as:
\begin{eqnarray}\label{eqn_tnfac}
t_n(T_e,T_0,N_e)&=& \frac{\Delta \nu_{rad}}{\Delta \nu_{col}}, \nonumber\\
&=& \left[\frac{6.096\times10^{-17}}{10^{a(T_e)}} \right] \left( \frac{T_0}{N_e} \right) n^{5.8-\gamma_c(T_e)}.
\end{eqnarray}
\noindent Note that the dependence on electron temperature is contained within the fitting coefficients, $a$ and $\gamma_c$. For $T_e=100~\mathrm{K}$,
we find $t_n\approx 5.82\times10^{-7} \sqrt{n} \left( T_0/N_e\right)$. In Figure~\ref{fig_tnfac}, we show $t_n$ as a function of electron density
for $T_0=1000~\mathrm{K}$. For a given electron temperature and density, the influence of an external radiation field is larger for higher levels
since $t_n \propto n^{5.8-\gamma_c}$ and $\gamma_c<5.8$ (see Appendix). For a given density, the influence of the radiation field on the line width
is larger at higher electron temperatures. For the typical conditions of the CNM, i. e. at $T_e=100~\mathrm{K}$ and $N_e=0.02~\mathrm{cm^{-3}}$, the value of $t_n\approx1$ and both radiation field
and electron density affect the line width in similar amounts.

\section{Method}\label{section_results}

\begin{figure}[!ht]
\includegraphics[width=1.\columnwidth]{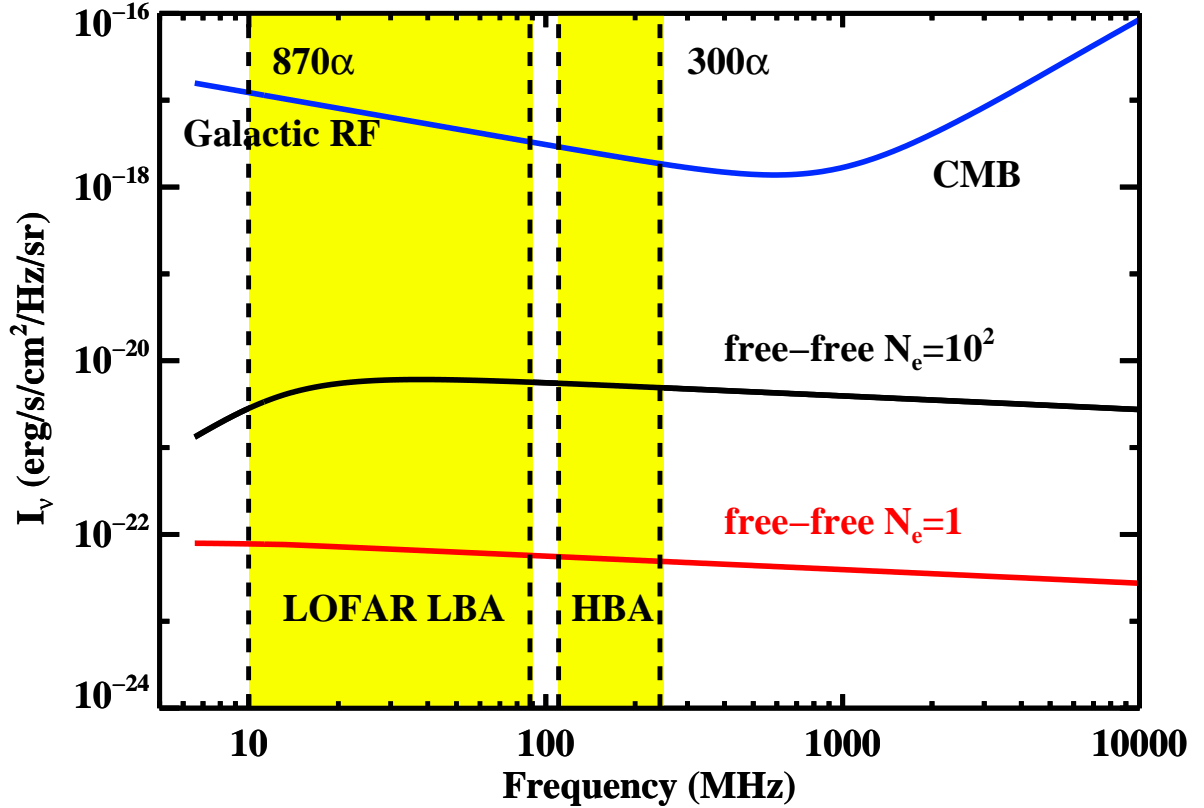}
\caption{A comparison between the continuum radiation fields. The galactic synchrotron radiation field dominates over the free-free cloud continuum
at $T_e=100~\mathrm{K}$. Therefore, the strong background approximation is valid for the low temperature cases considered in this analysis. The yellow zone marks
the range in frequency observable by LOFAR.\label{fig_contdiff}}
\end{figure}

In order to study the radiative transfer effects on the lines we use the method outlined in Paper~I to compute the departure coefficients for different electron
temperatures, densities and considering an external radiation field. The cosmic microwave radiation field (CMB) and the Galactic synchrotron power law radiation
field spectra are included. We represent the cosmic microwave radiation field (CMB) by a 3~K blackbody and the galactic radiation field by a power law [$I_0(\nu)=T_0(\nu/\nu_0)^{\alpha_{pl}}$]
with $T_0=1000~\mathrm{K}$ at a frequency $\nu_0=100~\mathrm{MHz}$ and $\alpha_{pl}=-2.6$ \citep{landecker1970, bennett2003}. In the Galactic plane, the Galactic
radiation field can be much larger than 1000~K at 100 MHz \citep{haslam1982}.
At frequencies higher than 1 GHz, corresponding to $\mathrm{C}n\alpha$ transitions from levels with $n<200$, the background continuum is dominated by the CMB
(Figure~\ref{fig_contdiff}). At even higher frequencies, the background continuum can be dominated by dust and free-free emission, which are strongly dependent
on the local conditions of the cloud and its position in the Galaxy. For simplicity, we focus our study on levels with $n>200$.

Departure coefficients were computed for $T_e=20,~50,~100~\mathrm{and}~200~\mathrm{K}$ and electron densities in the range $10^{-2}~\mathrm{to}~1~\mathrm{cm^{-3}}$. Once
the departure coefficients were obtained, we computed the corresponding optical depths assuming a fixed length along the line of sight of 1~pc 
from the usually adopted approximated optical depth solution to the radiative transfer problem (Equation \ref{eqn_integlinetocont}).
The value of 1~pc corresponds to emission measures in the range of $EM_\mathrm{C+}=10^{-4}~\mathrm{to}~1~\mathrm{cm^{-6}~pc}$.
Our calculations assume a homogeneous density distribution in a cloud and should be taken as illustrative since it is well known that inhomogeneities
exist in most clouds. The fixed length of 1~pc corresponds to column densities of $10^{18}~\mathrm{to}~10^{21}~\mathrm{cm^{-2}}$, with the adopted density range.
Diffuse clouds show a power-law distribution function in HI column density with a median column density of $0.76\times10^{20}~\mathrm{cm^{-2}}$ \citep{heilesandtroland2003b}.
Reddening studies are weighted to somewhat large clouds and the standard ``Spitzer'' type cloud \citep{spitzer1978} corresponds to a column density of $3.6\times10^{20}~\mathrm{cm^{-2}}$. 
Local HI complexes associated with molecular clouds have $\mathcal{N}_\mathrm{H}\approx10^{21}~\mathrm{cm^{-2}}$.

\section{Results}
\subsection{Line widths}

\begin{figure}[!ht]
\includegraphics[width=1.0\columnwidth, angle=0]{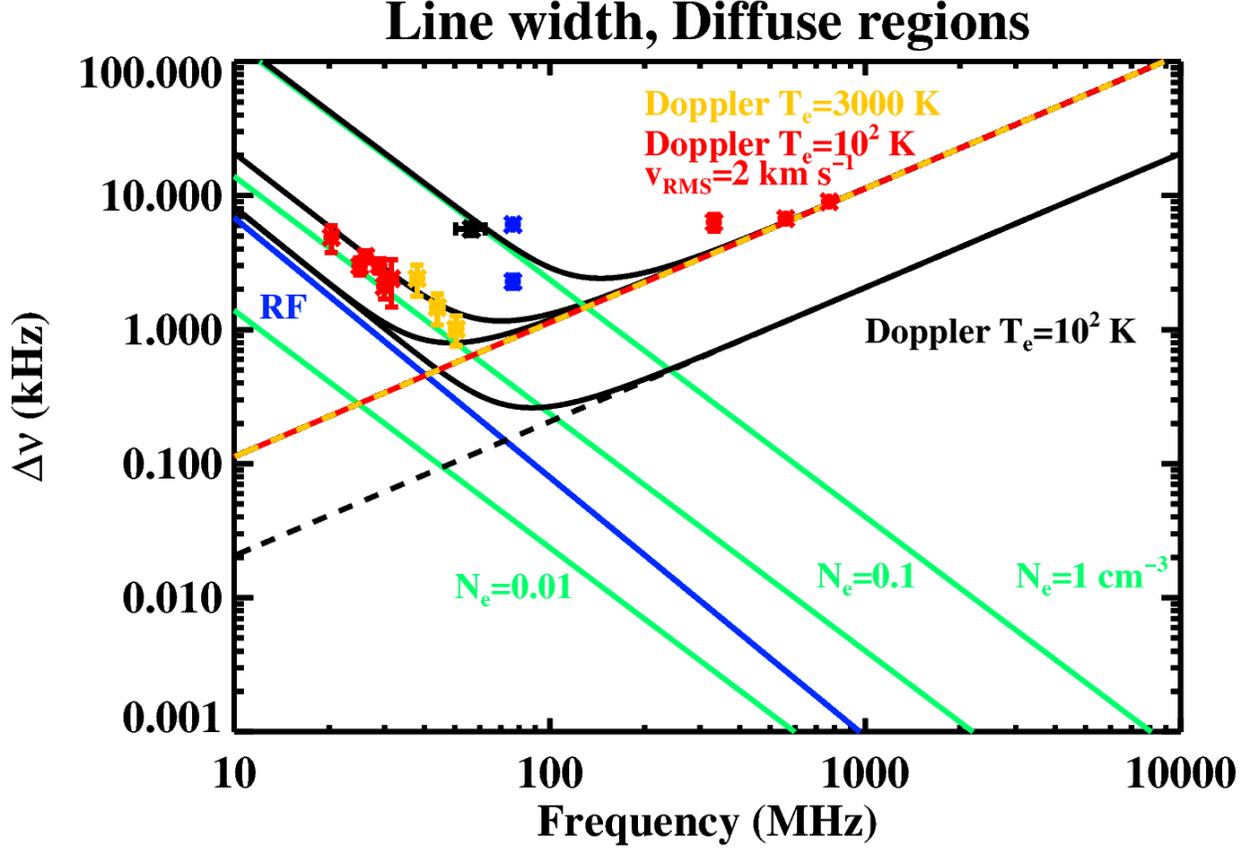}
\caption{A comparison between broadening produced by the Galactic radiation field (blue line), collisional broadening at
$N_e=1,~0.1~\mathrm{and}~0.01~\mathrm{cm^{-3}}$ (green lines) and thermal (Doppler) broadening at 100~K (black dashed line). The red and yellow curves correspond to a turbulent Doppler parameter $<v_{RMS}>^2=2~\mathrm{km~s^{-1}}$ and $T_e=300~\mathrm{K}$, respectively. We include data for Cas~A \citep{payne1994, kantharia1998} as red points, Cyg~A \citep{oonk2014} as yellow points, regions for the inner galaxy \citep{erickson1995} as blue points and data for M82 \citep{morabito2014} as a black point.
\label{fig_widthdiff}}
\end{figure}

We begin our discussion with the results for the line widths. We show the line widths for our diffuse clouds models in Figure~\ref{fig_widthdiff}.
At high frequencies (low $n$), the Gaussian (Doppler) core of the line dominates the line profile in frequency space and the line width increases with
frequency. At low frequencies (high $n$), on the other hand, the Lorentzian profile dominates --either because of collisional or radiation broadening--
and the line width decreases with increasing frequency. In order to guide the discussion we have included observed line widths for $\mathrm{C}n\alpha$
transitions for Cas~A \citep{kantharia1998, payne1994}, Cyg~A \citep{oonk2014} and M82 \citep{morabito2014} .

When the Doppler core dominates, CRRLs observations provide both an upper limit on the gas temperature and an upper limit on the turbulent velocity
of the diffuse ISM (cf. Equations~\ref{eqn_radbroaden} and \ref{eqn_colbroaden}). For typical parameters of the turbulent ISM ($1~\mathrm{km~s^{-1}}$), turbulence dominates over
thermal velocities when $T_e\lesssim 700~\mathrm{K}$.

Radiation broadening and collisional broadening show a very similar dependence on $n$ and it is difficult to disentangle these two processes
from CRRLs observations. For the Galactic radiation field (i.e. synchrotron spectrum with $T_0=1000~\mathrm{K}~\mathrm{at}~100~\mathrm{MHz}$), the two
processes contribute equally to the line width at a density $N_e\approx0.03~\mathrm{cm^{-3}}$ (Figure~\ref{fig_widthdiff}). Low frequency observations can,
thus, provide an upper limit on the density and radiation field. As illustrated in Figure~\ref{fig_widthdiff}, the transition from a Doppler to a
Lorentzian broadened line is quite rapid (in frequency space) but the actual value of $n$ where it occurs depends on the physical conditions of the cloud (i.e. $T_e,~N_e,~T_0~\mathrm{and}~\langle v_{RMS} \rangle$).

In Figure~\ref{fig_widthdiff} we can see that the RRLs from Cas~A and Cyg~A fall in a region of the diagram corresponding to densities lower than about
$0.1~\mathrm{cm^{-3}}$ and the detection in M82 corresponds to either higher densities, to a much stronger radiation field or to the blending
of multiple broad components. From observations at high frequencies, it is known that the lines observed towards Cas~A are the result of three
components at different velocities in the Perseus and Orion arms. Therefore, the physical parameters obtained from line widths should be taken as upper limits.

\begin{figure}[!ht]
\includegraphics[width=1\columnwidth, angle=90]{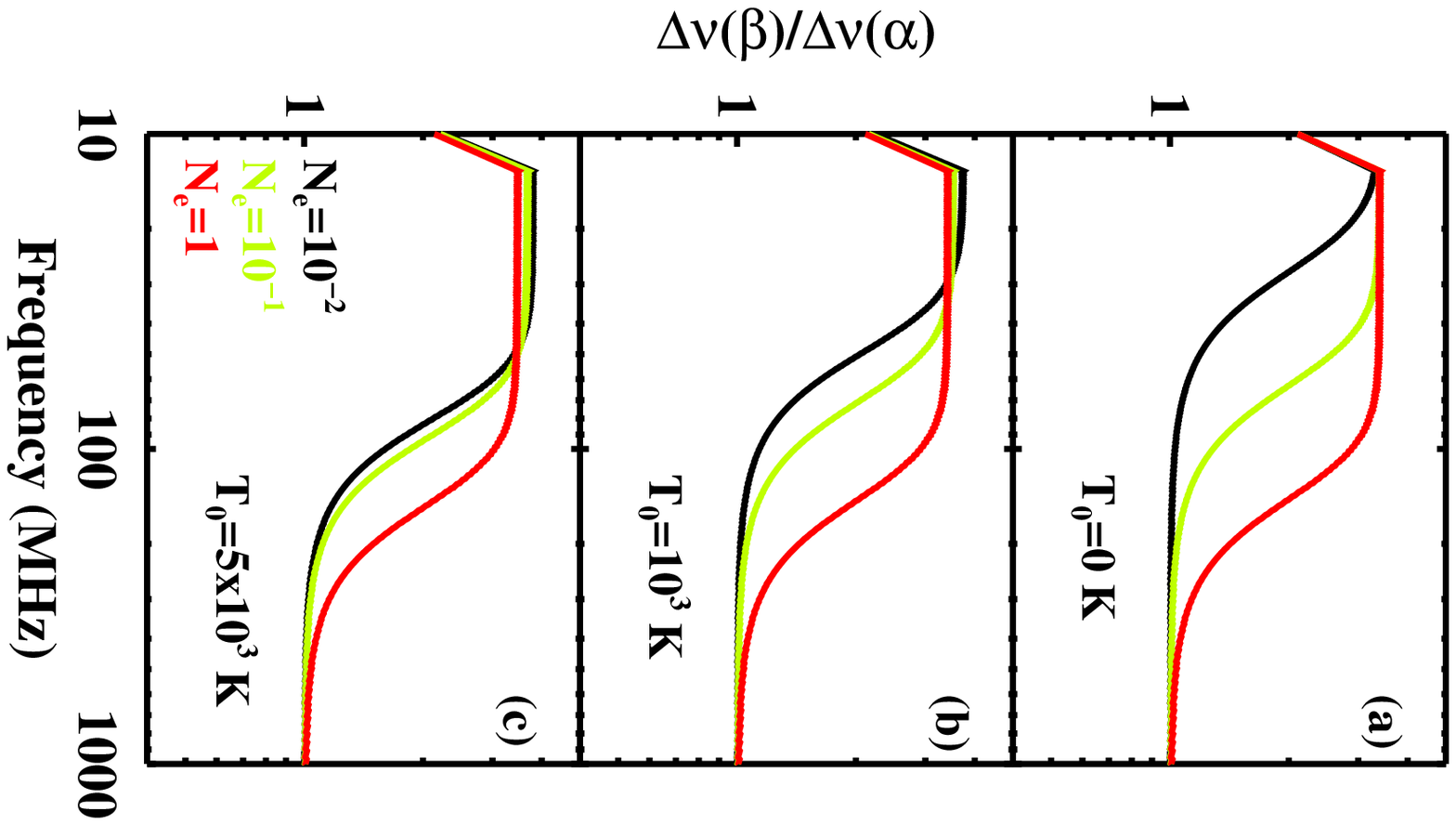}
\caption{$\alpha~\mathrm{and}~\beta$ line width transitions for diffuse regions as a function of frequency for different power law radiation fields. a) Without an external radiation field; b) a power law radiation field with $T_0=1000~\mathrm{K}$ and c) as b) for $T_0=5000~\mathrm{K}$. The line widths correspond to electron densities of $N_e=1,~0.1,~\mathrm{and}~0.01~\mathrm{cm^{-3}}$ (red, green and black lines).
\label{fig_widthdiffbetagamma}}
\end{figure}

When the line profile is dominated by the Doppler core, the ratio of the $\beta$ to $\alpha$ line width is unity. However, radiation or collisional broadening
affects the $\mathrm{C}n\alpha$ and $\mathrm{C}n\beta$ lines differently as at the same frequency $\mathrm{C}n\alpha$ and $\mathrm{C}n\beta$ lines originate
from different $n$ levels. In Figure \ref{fig_widthdiffbetagamma}, we show the ratio $\Delta \nu(\beta)/\Delta \nu(\alpha)$. We notice that, when radiation broadening dominates
the line width, this ratio goes to a constant value, independent of the background temperature. From the radiation broadening formula (Equation \ref{eqn_radbroaden}) we see that
$\Delta \nu(\beta)/\Delta \nu(\alpha)=(n_\beta/n_\alpha)^{-3\alpha_{pl}-2}$ and, for a power law $\alpha_{pl}=-2.6$, the ratio approaches $\Delta \nu(\beta)/\Delta \nu(\alpha)=3.8$
as $n$ increases. At high electron densities, collisional processes dominate the broadening of the lines. From Equation \ref{eqn_colbroaden} the $\Delta \nu(\beta)/\Delta \nu(\alpha)$
ratio tends to a constant value of $(n_\beta/n_\alpha)^{\gamma_c}=1.26^{\gamma_c}$. There is a temperature dependence in the exponent $\gamma_c$ and, for electron
temperatures less than 1000 K, we find that $\Delta \nu(\beta)/\Delta \nu(\alpha)\approx3.1-3.6$ (see Table~\ref{table_col}); similar to the radiation broadening case.

\subsection{Integrated line to continuum ratio}

As discussed in section~\ref{section_radtransf}, the line to continuum ratio of CRRL is often solved approximately, using equation~\eqref{eqn_integlinetocont}.
In this subsection, we will discuss when this approximation is justified.
In this, we have to recognize that, under the conditions of the diffuse ISM, recombining carbon atoms are not in LTE (Paper~I). Indeed, electrons can
recombine to high levels due to dielectronic recombination, thus increasing the population in comparison to the LTE values. This increase in the level
population leads to an increase in the values of the $b_n\beta_n$ coefficients in Equation~\ref{eqn_integlinetocont} and, consequently, to an increase
in the optical depth of the lines.

\begin{figure}[!ht]
\includegraphics[width=0.5\columnwidth]{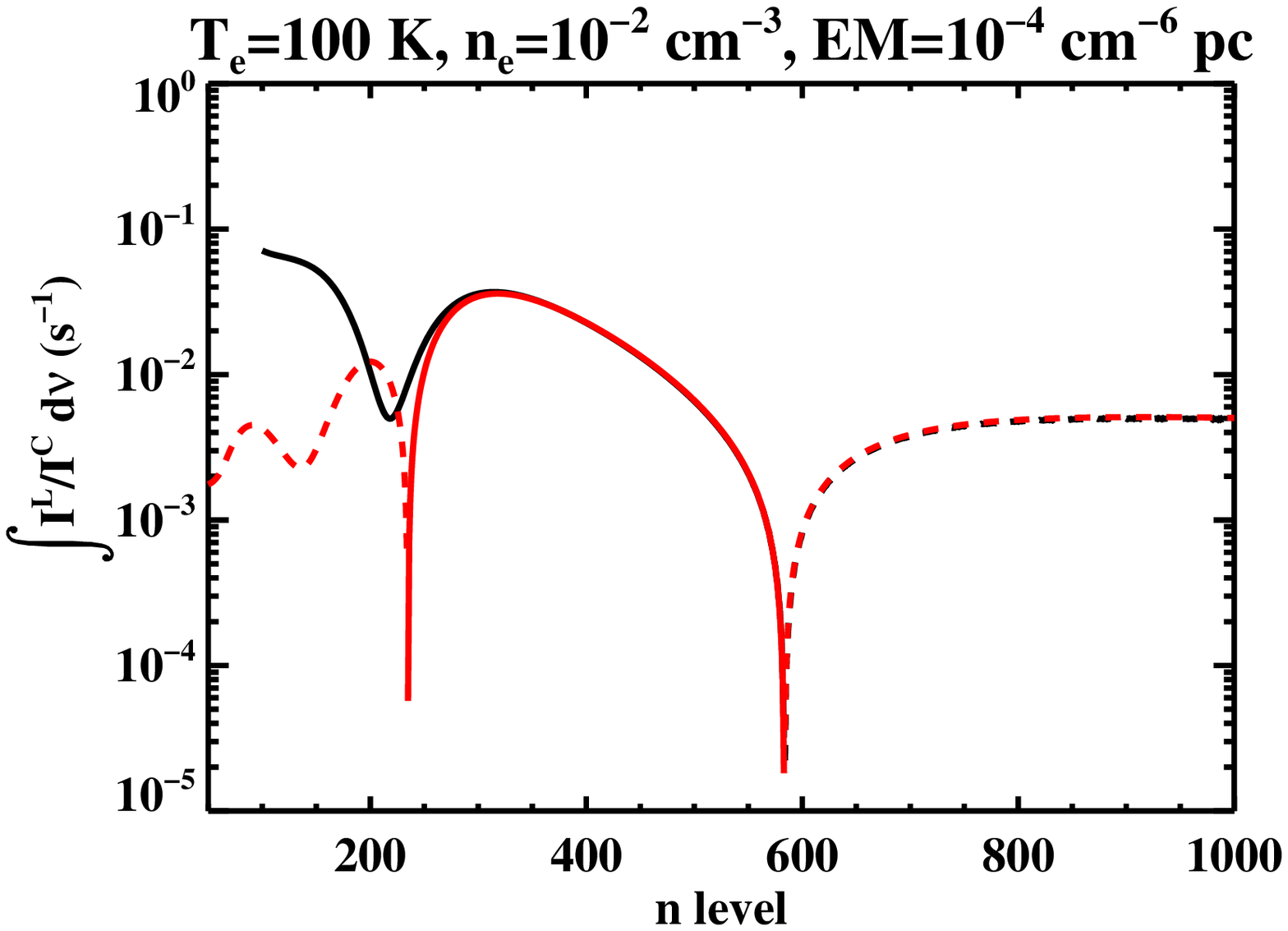}
\includegraphics[width=0.5\columnwidth]{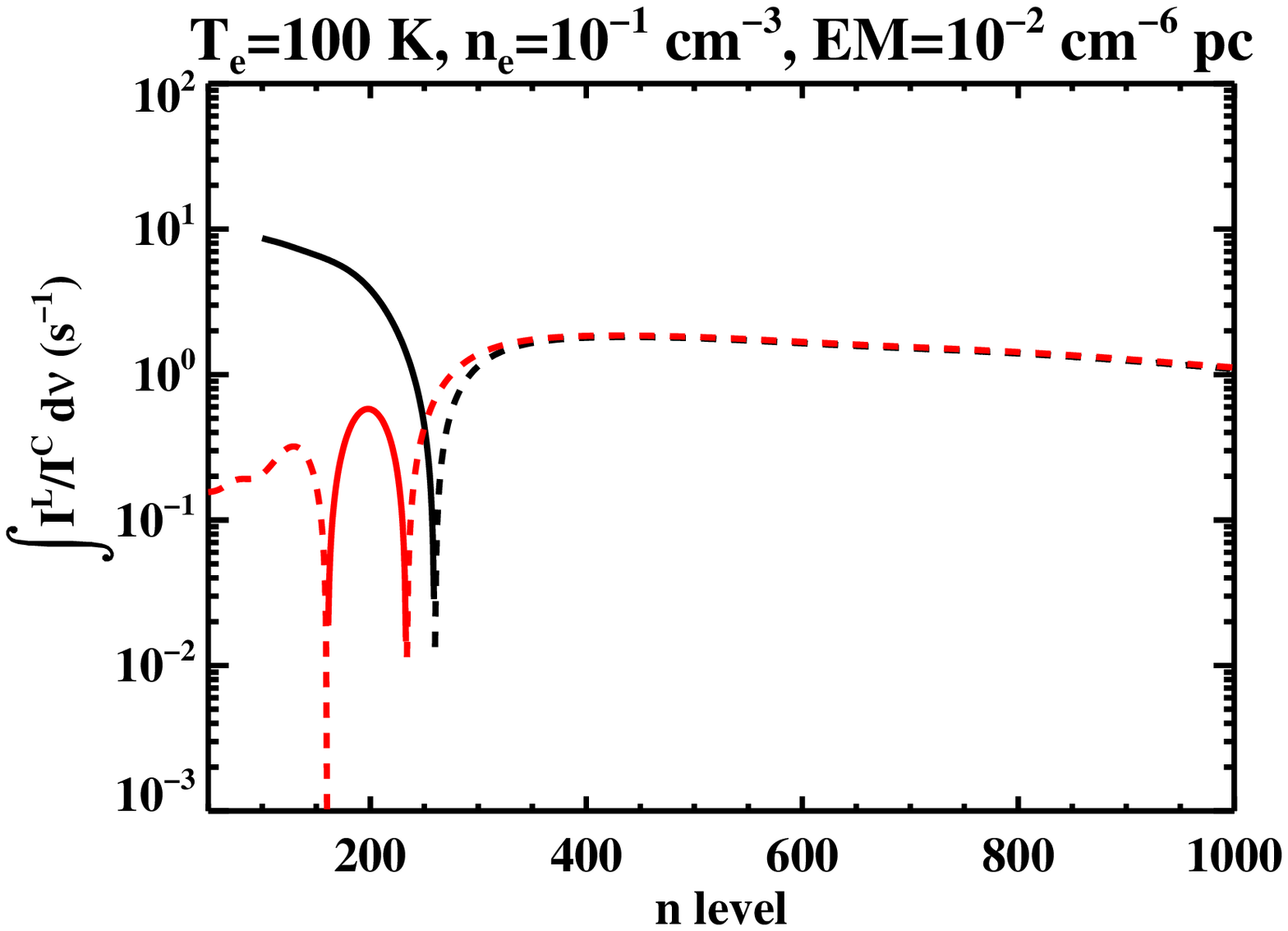}
\caption{The line-to-continuum ratio of CRRL as a function of principal quantum number for $T_e=100~\mathrm{K}$ and $N_e=0.01~\mathrm{and}~0.1~\mathrm{cm^{-3}}$
(left and right panels, respectively). The values were computed from the radiative transfer solution (Equation \ref{eq_linecontgeneral}) and the galactic
radiation field as a background. Black lines correspond to the result of solving the equation of radiative transfer while red lines correspond to the approximation
expression given in Equation~\ref{eqn_integlinetocont}. At levels larger than $n\gtrsim250$, the differences between the approximation (dashed) and
the radiative transfer solution (solid) are minor. \label{fig_integtau}}
\end{figure}

In Figure~\ref{fig_integtau} we show the integrated line to continuum ratio as a function of level $n$ for $T_e=100~\mathrm{K}$. We compare
the values obtained using the approximated expression given in Equation~\ref{eqn_integlinetocont} (red lines) and by solving the radiative
transfer equation (Equation~\ref{eq_linecontgeneral}, black lines). The agreement between the two approaches is good for levels $n\gtrsim250$, since
at these high levels the approximations that lead to Equation~\ref{eqn_integlinetocont} are valid.
For levels lower than $n\approx250$, differences appear. In particular, at low electron densities ($N_e\approx0.01~\mathrm{cm^{-3}}$)
results using Equation~\ref{eqn_integlinetocont} show lines in absorption while the results derived from solving the radiative transfer equation predict
lines in emission. The difference between the two approaches can be understood in terms of the excitation temperature (see Appendix). As can be seen in Figure~\ref{fig_integtau2},
the red zones correspond to low $n$ levels, where the excitation temperature is higher than the background continuum temperature and the lines appear in
emission (despite the $\beta_n$ being positive). At higher $n$ values, $\beta_n < 0$ (yellow zones) and the excitation temperature is negative reflecting an
inversion in the level population, consequently, lines appear in emission. While there is an inversion of the level population the line optical depths
are too low ($\tau_l\sim 10^{-3}$) to produce a maser (cf. Equation~\ref{eqn_integlinetocontx}). At even higher levels (blue zones in Figure~\ref{fig_integtau2}),
the excitation temperature is less than the background continuum temperature and the lines are in absorption. As the electron density increases,
dielectronic recombination is less efficient and the levels for which $\beta_n$ is negative shift to lower $n$ values, resembling the values for hydrogenic
level population (\citealt{hummer1987}, Paper~I). Furthermore, for high quantum numbers and high densities, $\beta_n=1$ and the excitation temperature is
equal to the electron temperature of the gas.

\begin{figure}[!ht]
\includegraphics[width=0.5\columnwidth]{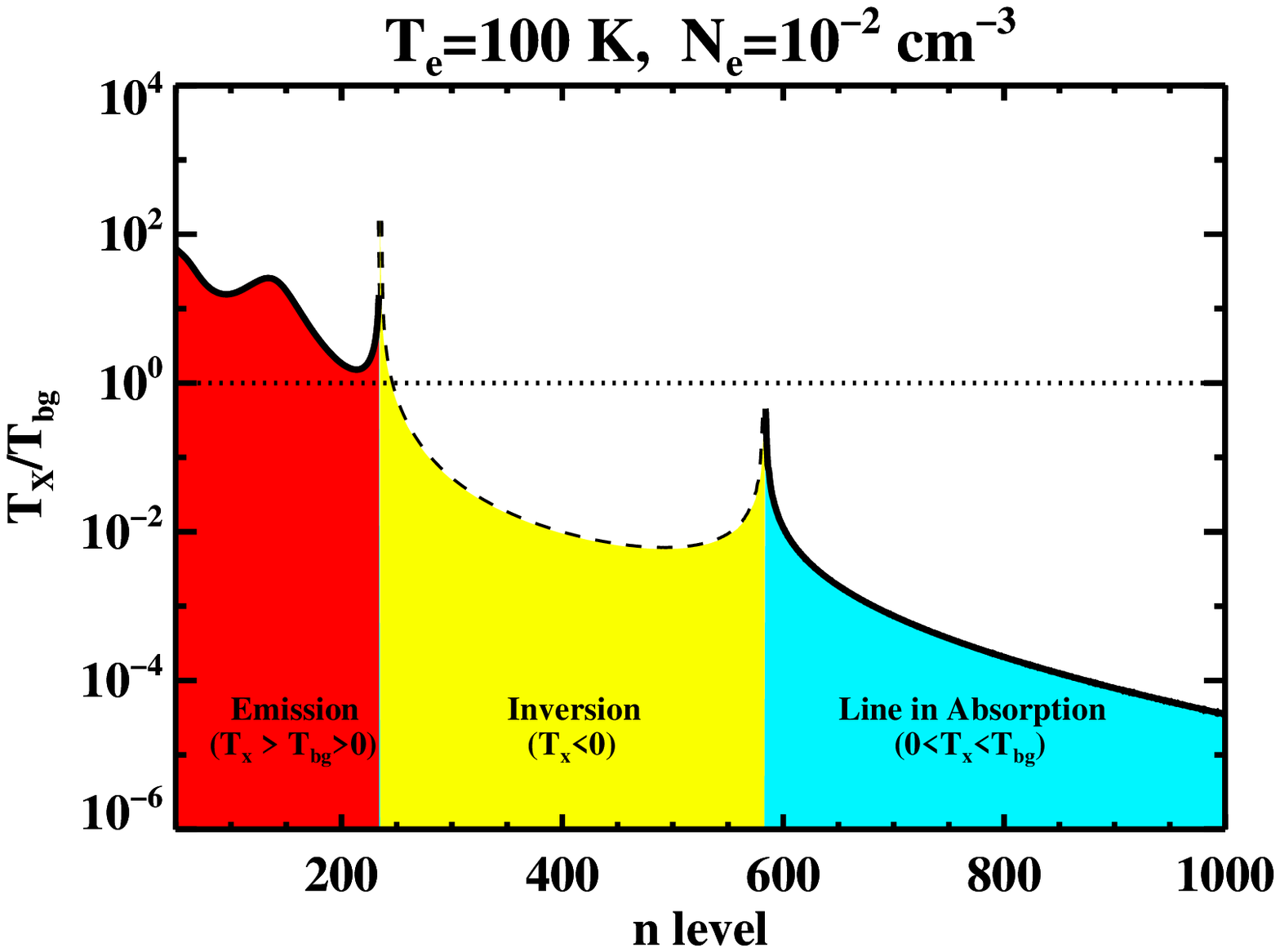}
\includegraphics[width=0.5\columnwidth]{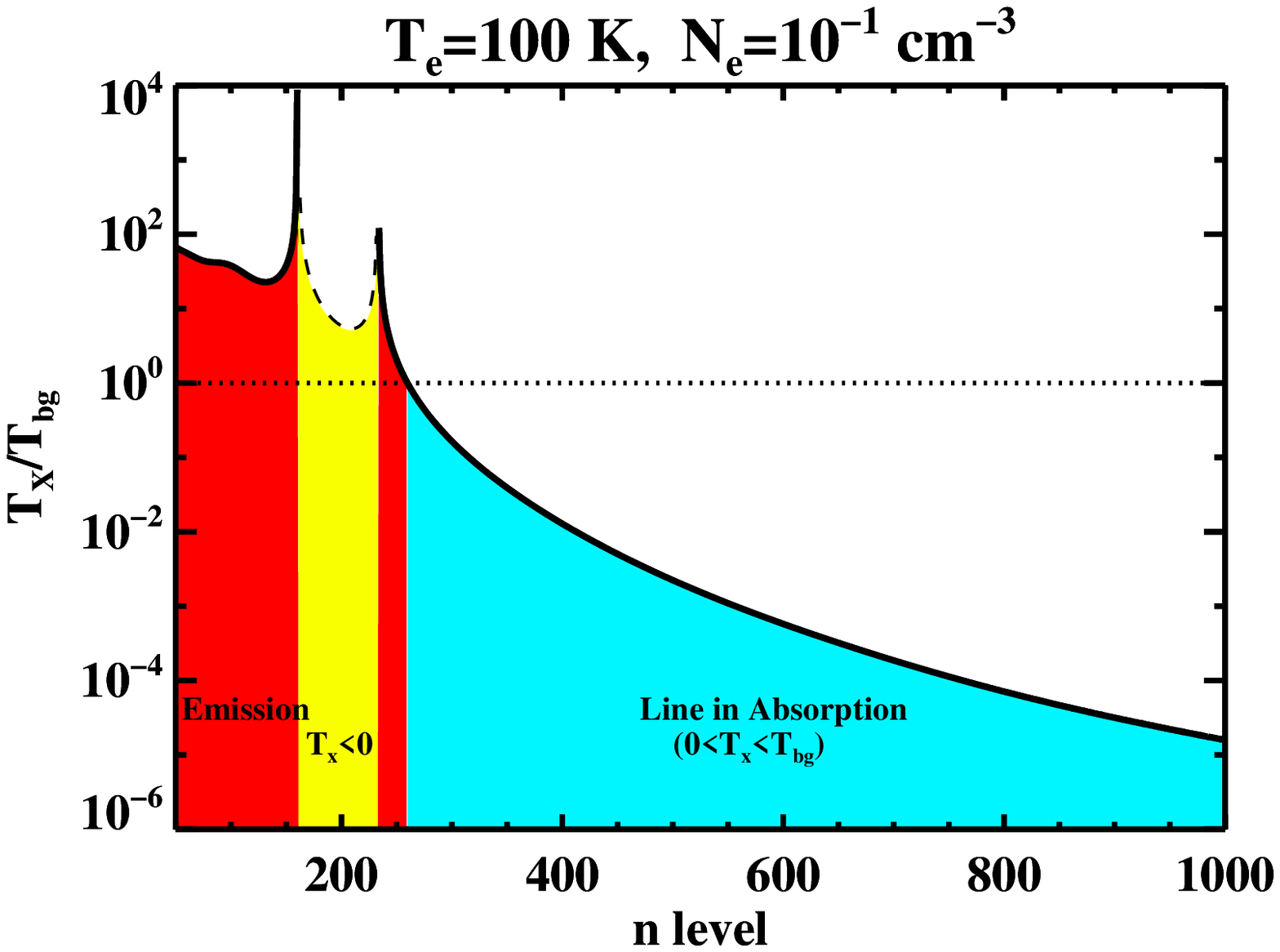}
\caption{Ratio of the excitation to background temperature ($T_X/T_{bg}$). Lines are in emission in the red zone since $T_X > T_{bg}$ and in the yellow zone
due to an inversion on the level population and $T_X< 0$. Lines appear in absorption in the light blue zone since the background temperature is (much) larger
than the excitation temperature.\label{fig_integtau2}}
\end{figure}

From this analysis, we conclude that Equation~\ref{eqn_integlinetocont} is valid for high ($n\gtrsim250$) quantum numbers and the ratio of two lines depends
only on the temperature and electron density of the cloud through the departure coefficients.
In Figure~\ref{fig_integtau3}, we demonstrate this by showing the integrated line to continuum ratio of $\mathrm{C}n\alpha$ as a function of
quantum number normalized to the level 500 (similar results can be obtained by using other $n$ levels). The normalized ratio becomes smaller
for high densities owing to the fact that $b_n\beta_n$ values change little with $n$ as the levels are closer to equilibrium. As the electron
density decreases, dielectronic recombination is more efficient in overpopulating intermediate levels (Paper~I) producing large changes in the values
of the ratios. 

\subsection{CRRLs as diagnostic tools for the physical conditions of the ISM}

\subsubsection{Line Ratios}
\begin{figure}[!ht]
\includegraphics[width=0.9\columnwidth, angle=90]{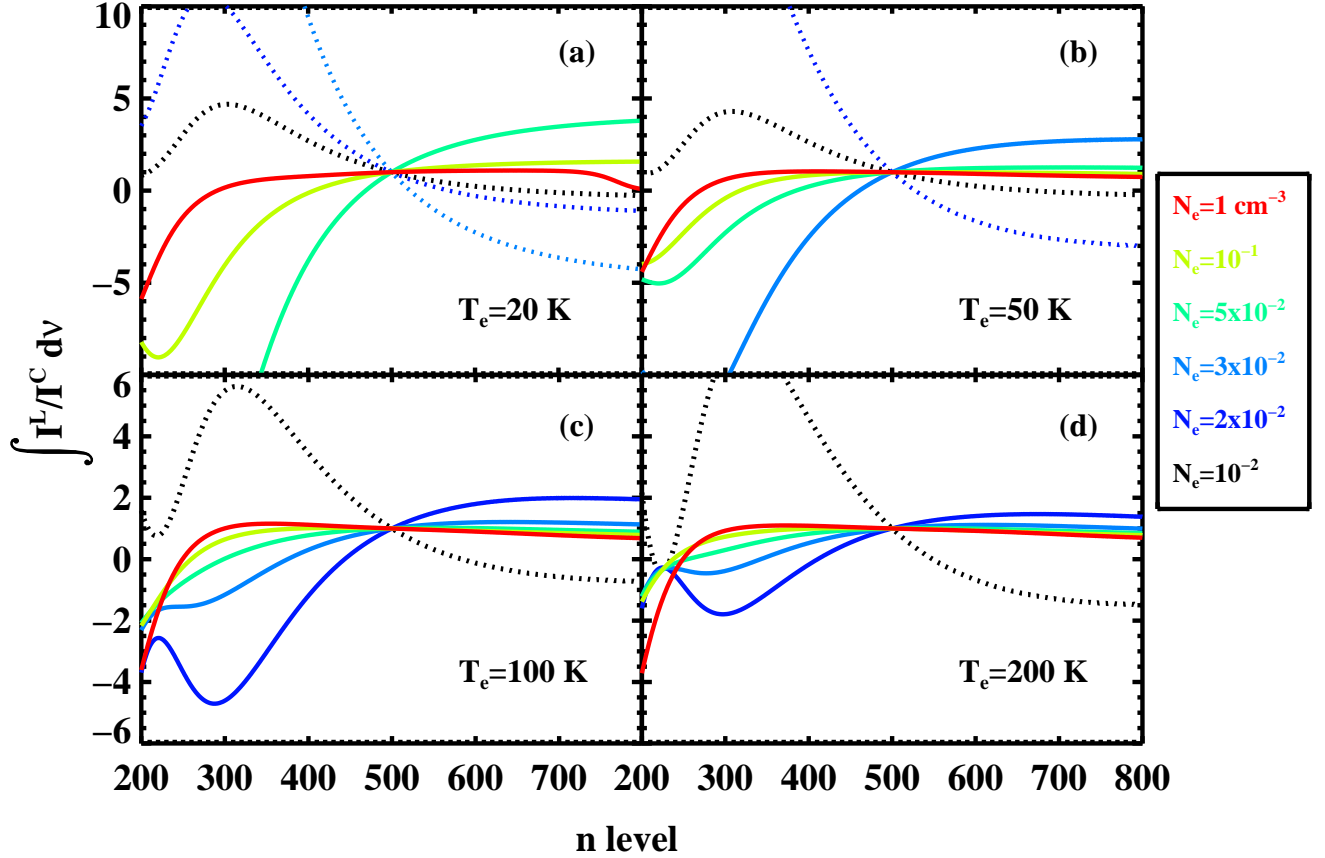}
\caption{Integrated line to continuum ratio normalized to the value at the level $n=500$ for $T_e=20,~50,~100,~\mathrm{and}~200~\mathrm{K}$. Dotted
lines indicate that the $\mathrm{C}500\alpha$ line is in emission. The values have been computed considering radiative transfer effects (Equation \ref{eq_linecontgeneral}).\label{fig_integtau3}}
\end{figure}

\begin{figure}[!ht]
\includegraphics[width=1\columnwidth, angle=90]{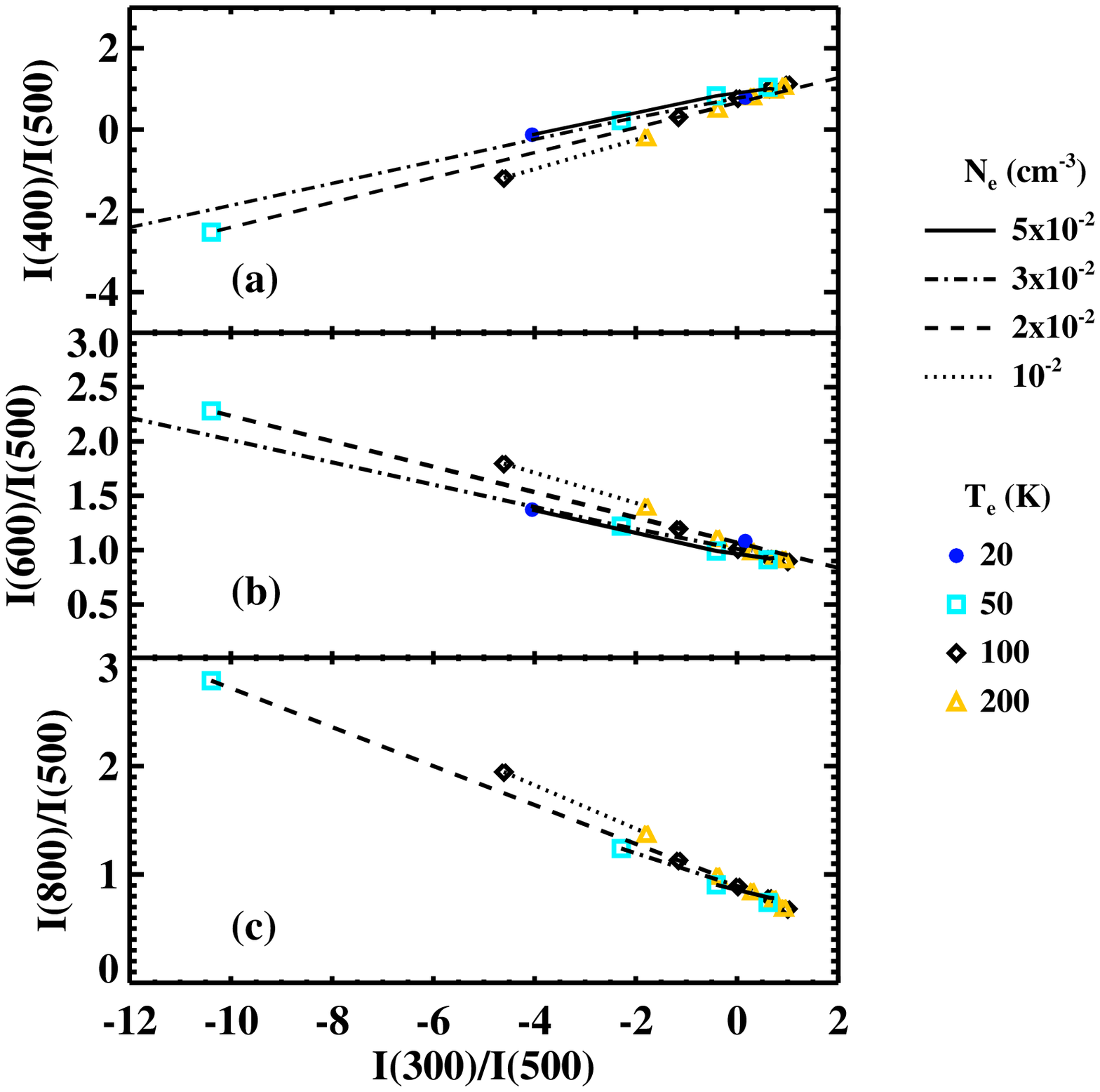}
\caption{Example ratio diagnostic plots for different electron temperatures and densities. Cyan points are at $T_e=50~\mathrm{K}$, black points for $T_e=100~\mathrm{K}$ and orange points for
$T_e=200~\mathrm{K}$. Different densities are joined by: dotted lines ($N_e=10^{-2}~\mathrm{cm^{-3}}$), dashed lines ($N_e=2\times10^{-2}~\mathrm{cm^{-3}}$),
dashed-dotted lines ($N_e=3\times10^{-2}~\mathrm{cm^{-3}}$) and continuous lines ($N_e=5\times10^{-2}~\mathrm{cm^{-3}}$). 
(a) Ratio of the integrated line to continuum for levels 400 and 500 vs. 300 to 500 ratio. (b) Ratio of the integrated line to continuum for levels 600 and
500 vs. 300 to 500 ratio. (c) Ratio of the integrated line to continuum for levels 800 and 500 vs. 300 to 500 ratio.
\label{fig_integtauratio}}
\end{figure}

We have already discussed the use of the line width to constrain the properties of the emitting/absorbing gas. As figure~\ref{fig_integtau3} illustrates, line ratios are
very sensitive to the physical conditions in the gas. Moreover, the use of line ratios ``cancels out'' the dependence on the emission measure. Here, we demonstrate the use of line
ratios involving widely different $n$'s as diagnostic tools in ``ratio vs. ratio'' plots. As an example, we show three line ratios in Figure~\ref{fig_integtauratio},
normalized to $n=500$. The lines are chosen to sample the full frequency range of LOFAR and the different regimes (collisional, radiative) characteristic for CRRLs.
The $n=300,~400,~500$ lines are a particularly good probe of electron density for regions with temperature less than about $100~\mathrm{K}$. The use of the $n=500$
level does not affect our results and other levels (e.g. $n=600~\mathrm{or}~800$) may be used for computing the ratios. We note that in a limited but relevant
electron density range ($N_e\sim1-5\times10^{-2}~\mathrm{cm^{-3}}$), these lines can be good tracers of temperature. At higher densities, the departure coefficients
approach unity and the ratios tend to group in a small region of the plot and the use of the ratios as probes of temperature requires measurements with high signal
to noise ratio to derive physical conditions from the observations.

\subsubsection{The Transition from Absorption to Emission}

\begin{figure}[!ht]
\includegraphics[width=1\columnwidth]{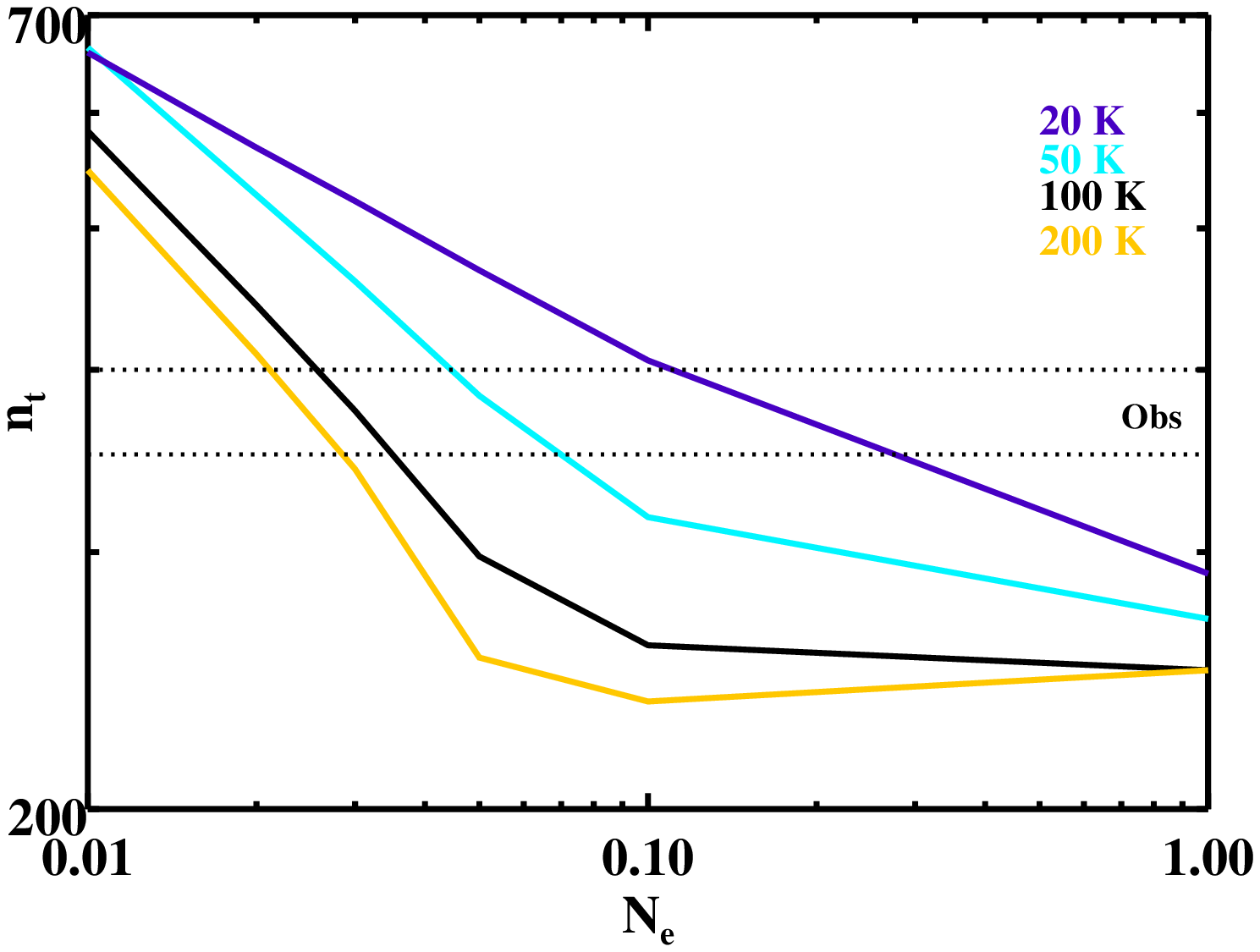}
\caption{Level where lines transition from emission to absorption ($n_t$) as a function of electron density ($N_e$)
for $T_e=50,~100~\mathrm{and}~200~\mathrm{K}$. The horizontal dashed lines mark the limits as suggested by observations of CRRLs in the Galaxy.
\label{fig_ntvsne}}
\end{figure}

In Paper~I, we discussed the use of the level where lines transition from emission to absorption ($n_t$) as a constraint on the density of
a cloud (Figure~\ref{fig_ntvsne}). The limited observations in the Galactic plane \citep{erickson1995,kantharia2001} indicate that $400>n_t>350$ and $n_t$ depends
on both temperature and density. The transition level can be used to estimate the electron density for electron densities lower than about $10^{-1}~\mathrm{cm^{-3}}$.
For increasing electron density it becomes more difficult to constrain this quantity from the transition level alone.

\subsubsection{Line Ratios as a Function of $\Delta n$}

\begin{figure}[!ht]
\includegraphics[width=0.9\columnwidth,angle=0]{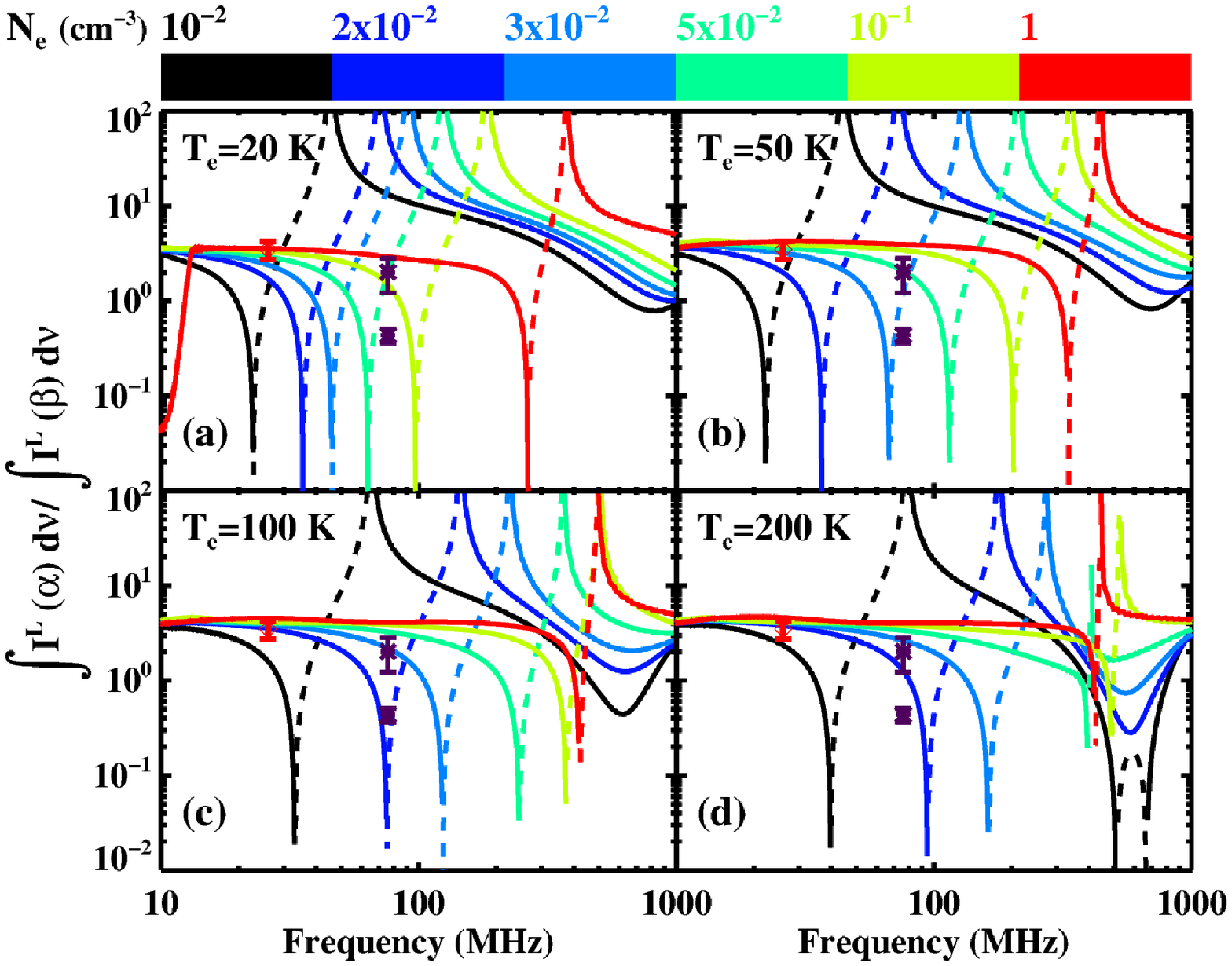}
\caption{Comparison between the integrated line to continuum $I(\alpha)/I(\beta)$ ratio as a function of frequency for different densities (colorbar);
dashed lines indicate that the ratio is negative, the color of the lines is the same as in Figure \ref{fig_integtau3}. The values 
for the ratios approach the LTE value of 3.6 at high $n$. Large differences can be observed for different densities because lines
observed at the same frequency correspond to different levels. We have included the data points for Cas~A from \citet{stepkin2007}
(red point) and for inner Galaxy from \citet{erickson1995} (dark blue points).\label{fig_alphabetagammaratios}}
\end{figure}

Combining observations of $\mathrm{C}n\alpha$ lines with $\mathrm{C}n\beta$ and $\mathrm{C}n\gamma$ lines can provide further constraints on
the physical parameters of the cloud. In Figure~\ref{fig_alphabetagammaratios} we show the $\alpha$-to-$\beta$ ratio of the integrated line to continuum ratio
as a function of frequency. Recall that $\mathrm{C}n\alpha~\mathrm{and}~\mathrm{C}n\beta$ lines observed at almost the same frequency probe very
different $n$ levels ($n_\alpha=1.26n_\beta$). Figure~\ref{fig_alphabetagammaratios} shows that both electron density and temperature are involved.
At high $n$ levels the $b_n \beta_n$ are approximately unity and the $\alpha$-to-$\beta$ approaches $M(1)/2 M(2)\approx0.1908/0.0526=3.627$
(Equation~\ref{eqn_integlinetocont}) making the ratio less useful to constrain temperature and electron density. However, even at high $n$,
this ratio does remain useful for investigating the radiation field incident upon the CRRL emitting gas.

\subsubsection{The CRRL/[CII] Ratio}

\begin{figure}[!ht]
\includegraphics[width=1\columnwidth]{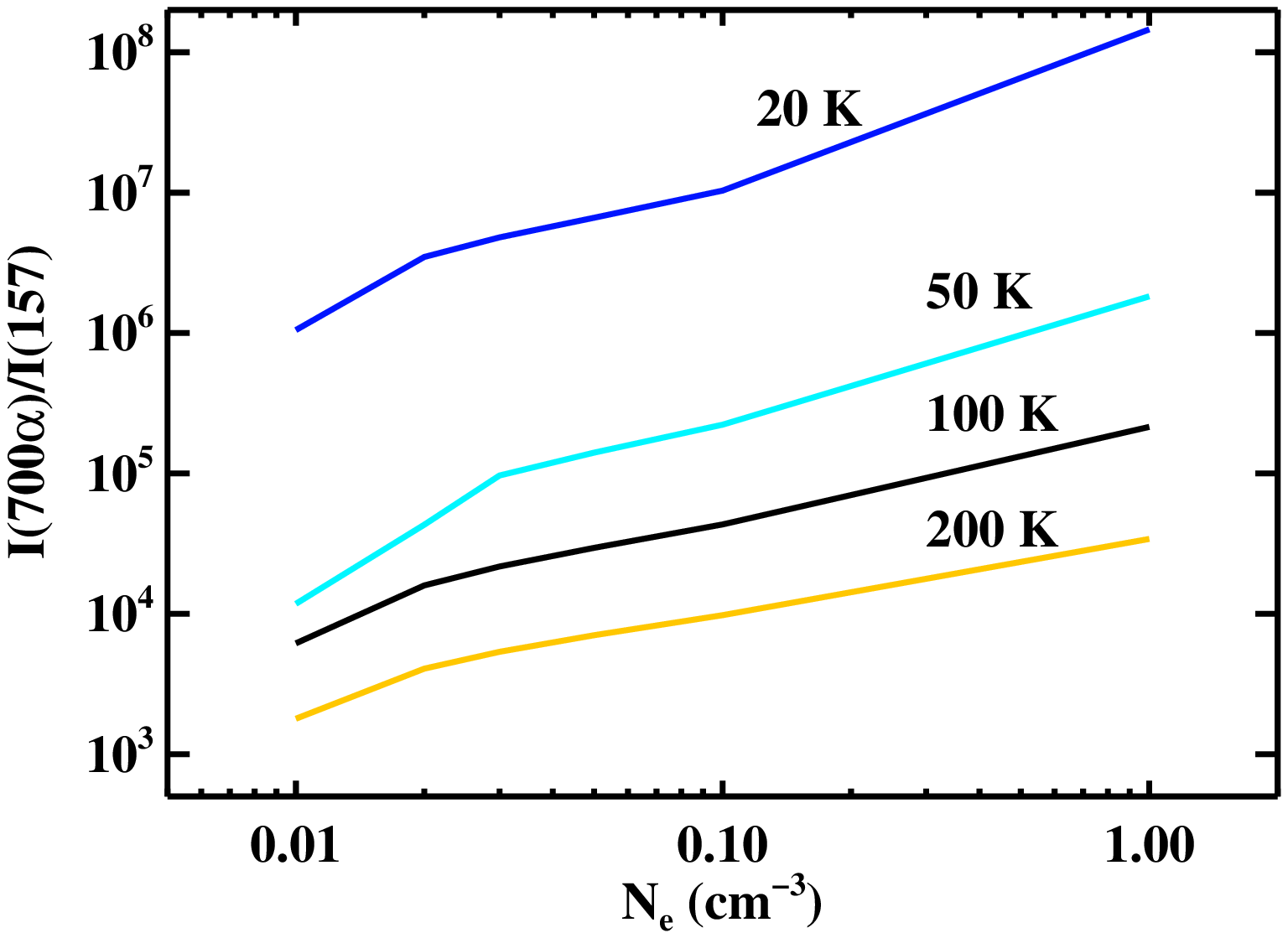}
\caption{Ratio of the line to continuum ratio, for $n=700$, to the [CII] $158~\mu m$ line as a function of density.
This example ratio shows how CRRL/[CII] can be used as a diagnostic plot to constrain electron density and temperature.\label{fig_diagplotdensity}}
\end{figure}

The [CII] 158~$\mu\mathrm{m}$ is the dominant cooling line of diffuse clouds and acts as a thermostat regulating the temperature \citep{hollenbach1999}.
In realistic models of the ISM of galaxies (e.g. \citealt{wolfire1995}), the photoelectric effect on polycyclic aromatic hydrocarbon molecules and very small
grains heats the gas and the cooling by the [CII] 158~$\mu\mathrm{m}$ line adjust to satisfy the energy balance. As the heating is a complicated function of
the physical conditions \citep{bakes1994}, models become very involved.
Here, we sidestep this issue and we calculate the [CII] 158 $\mu\mathrm{m}$ intensity as a function of $N_e$ and $T_e$ for a uniform cloud. The intensity scales with
the column density of carbon ions, $\mathcal{N}_\mathrm{C^+}$, and temperature. In contrast, the CRRLs scale with the emission measure
divided by $T_e^{5/2}$ (cf. Equation~\ref{eqn_integlinetocont}). Hence, the ratio of the CRRL to the 158 $\mu\mathrm{m}$ line shows a strong dependence on
temperature (and electron density), but for a constant density this ratio does not depend on column density. In Figure~\ref{fig_diagplotdensity}
we show the CRRL/[CII] ratio as a function of density for different temperatures. For the physical conditions relevant for diffuse clouds, the CRRL/[CII] ratio is
a powerful diagnostic tool. Moreover, as we demonstrate below, low frequency CRRLs are not expected to be observable at the typical temperatures
and densities of classical HII regions. We recognize that [CII] at 158 $\mu$m can be produced by the WIM. Nevertheless, we expect
the contribution from the WIM to the [CII] line to be $\sim 4\%$ in the general ISM\citep{pineda2013}. However, COBE observations of
the [NII] 205 $\mu$m line from the Milky Way (Bennet et al 1994) have demonstrated that [CII] 158 $\mu$m emission from the WIM may be
more important along some sight-lines (Heiles 1994).

\section{On the Observed Behavior of CRRLs}\label{section_examples}

\subsection{General considerations}

CRRLs have been observed towards two types of regions: high density PDRs and diffuse clouds \citep{gordon2009}.
In general, low frequency CRRLs are observed in absorption with values for the integrated line to continuum ratio in
the range of $1~\mathrm{to}~5~\mathrm{Hz}$ \footnote{We quote the integrated line to continuum ratio in units of Hz as opposed to $\mathrm{km~s^{-1}}$. }
and a peak line-to-continuum ratio of $\sim10^{-4}~\mathrm{to}~10^{-3}$ \citep{erickson1995,kantharia1998,roshi2002,oonk2014}.

In order to observe CRRLs, carbon atoms must be singly ionized. In HII regions, carbon is found in higher stages of
ionization and the gas is dense and warm. Hence, recombination lines of the type we study here are not expected to be strong.
n photodissociation regions of high density, carbon atoms transition from ionized to neutral and into molecular (CO) around a visual extinction $A_V\approx 4$~mag, depending on the density and UV field.
Assuming $A_V=\mathcal{N}_\mathrm{H}/1.9\times10^{21}~\mathrm{mag~cm^{-2}}$ we can estimate the maximum column density of carbon that can be expected for such
a transition region. Assuming that carbon is fully ionized and a carbon abundance of ${1.6\times 10^{-4}}$, we obtain
a column density of carbon of $1.2\times10^{18}~\mathrm{cm^{-2}}$.

As mentioned in Section \ref{section_radtransf}, CRRLs produced in clouds with high temperatures are faint due to the strong dependence
of the line-to-continuum ratio on temperature. Therefore, regions of low temperature are favored to be observed using low frequency
recombination lines. These two considerations (low $T_e$ and $N_e$) set a range of electron density and temperature for which CRRLs
are easier to detect. Specifically, consider a medium with two phases in pressure equilibrium. From Equation \ref{eqn_linetocontapp},
the optical depth ratio scales then with:
\begin{eqnarray}
\frac{\tau_1}{\tau_2}\propto \frac{N_{e,1}^2}{T_{e,1}^{5/2}} \frac{T_{e,2}^{5/2}}{N_{e,2}^2} \frac{(b_n\beta_n)_1}{(b_n\beta_n)_2}\frac{L_1}{L_2}\propto
\left(\frac{T_2}{T_1} \right)^{9/2} \frac{(b_n\beta_n)_1}{(b_n\beta_n)_2}\frac{L_1}{L_2}.
\end{eqnarray}
For parameters relevant for the CNM and WNM ($T_{e,1}=80~\mathrm{K},~T_{e,2}=8000~\mathrm{K}$, respectively, \citealt{tielens05}), we have then $\tau_1/\tau_2\sim10^9 (b_n\beta_n)_1/(b_n\beta_n)_2 L_1/L_2$.
Clearly, CRRLs will overwhelmingly originate in cold, diffuse clouds. Therefore, unlike 21 cm HI observations, analysis of CRRL observations
is not hampered by confusion of CNM and WNM components.

The fact that low frequency recombination lines are observed in absorption sets a lower limit on the density for the clouds where
CRRLs are produced. Our models show that at electron densities lower than $10^{-2}~\mathrm{cm^{-3}}$ and for temperatures lower
than 200~K low frequency CRRLs are in emission.

\subsection{Illustrative examples}

In this section we illustrate the power of our models to derive physical parameters from observations of CRRLs.
We selected observations towards Cas~A as, to our knowledge, the clouds towards Cas~A are the best studied using CRRLs.
We then expand this illustration, by using observations of two regions observed towards the Galactic Center from \citet{erickson1995}.

\subsubsection{Cas A}
\begin{figure}[ht]
\includegraphics[width=1\columnwidth]{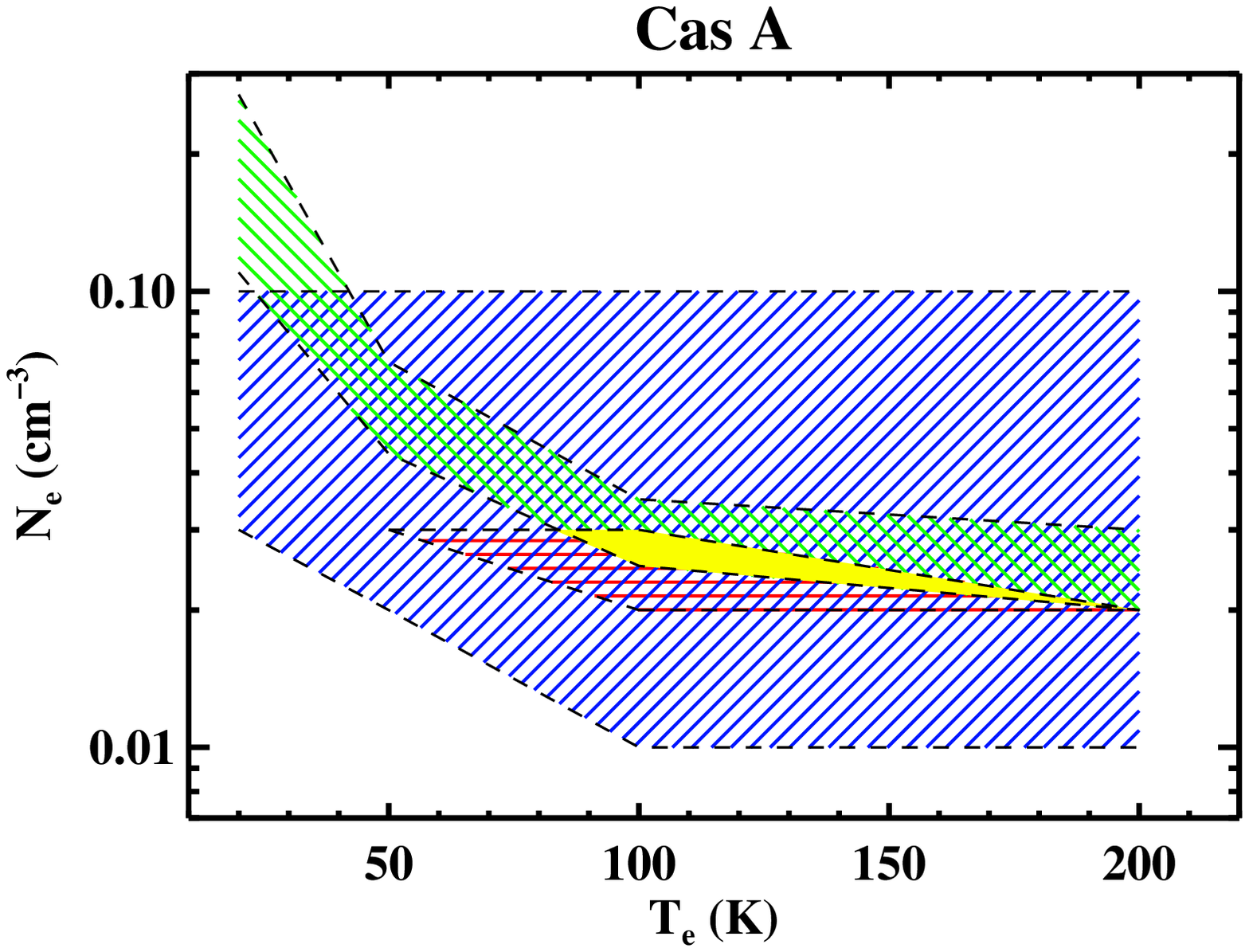}
\caption{Summary of the constraints for the $\mathrm{C}n\alpha$ and $\mathrm{C}n\beta$ transitions from \citet{stepkin2007} towards Cas~A. The blue zone
shows the region allowed by the integrated $\alpha$ to $\beta$ ratio constraints. The green zone is the region allowed from the $n_t$ constraints. The red zone is
the region allowed from the 600 to 500 ratio vs. 270 to 500. The yellow zone shows the overlap region from all the constraints. The electron density is well
constrained to be $2-3\times{10^{-2}}~\mathrm{cm^{-3}}$. The temperature is constrained to be within 80 and 200~K.\label{fig_contourcasa}}
\end{figure}

We begin our analysis with CRRLs detected towards Cas~A from the literature (e.g. \citealt{payne1994,kantharia1998, stepkin2007}).
In Figure \ref{fig_contourcasa}, we summarize the constraints from: the integrated line $\alpha$ to $\beta$ ratio as a blue zone
using the \citet{stepkin2007} data. The transition from emission to absorption ($350<n_t<400$) is shown as the green zone. The 600 to 500 ratio vs. 270 to 500 ratio is included
as the red zone\footnote{ We use the $n=270$ data from \citet{kantharia1998} and estimate the data at $n=600$ from \citet{payne1994} and analogous plots as in
Figure \ref{fig_integtauratio}.}. Finally, the yellow zone is the intersection of all the above mentioned zones. 

The line width does not provide much of an additional contraint. For Cas~A, with an observed line width of $6.7~\mathrm{kHz}$ at $\nu=560~\mathrm{MHz}$ \citep{kantharia1998}
the implied gas temperature would be $T_e=3000~\mathrm{K}$ and actually we expect that the line is dominated by turbulence with $<v_{RMS}>\approx 2~\mathrm{km~s^{-1}}$ (Figure~\ref{fig_widthdiff}).
Likewise, the Cas~A observations from \citet{payne1994,kantharia1998} are of little additional use as we arrive at $N_e\lesssim0.1~\mathrm{cm^{-3}}$ and $T_0\lesssim2000~\mathrm{K}$.

Perusing Figure~\ref{fig_contourcasa}, we realize that the $\alpha$ to $\beta$ line ratio does not provide strong constraints due to the frequency at which the lines were
observed as all the models converge to the high density limit (Figure \ref{fig_alphabetagammaratios}). The transition level from emission to absorption ($n_t$) restricts
the allowed models to an area in the $N_e~\mathrm{vs}~T_e$ plane. However, at low temperatures ($T\lesssim 50$ K), the constraining power of $n_t$ is limited.
The ``ratio vs. ratio'' plots can be quite useful in constraining both the electron density and the temperature of the line producing cloud, as we have illustrated here.

The results of our models show that the properties of the cloud are well restricted in density ($N_e=2-3\times10^{-2}~\mathrm{cm^{-3}}$) --  corresponding to H-densities of
$\sim100-200~\mathrm{cm^{-3}}$ -- but somewhat less in temperature ($T_e=80-200~\mathrm{K}$). We emphasize, though, that these results are ill-defined averages as the CRRLs
towards Cas A are known to be produced in multiple velocity components which are blended together. In addition, preliminary analysis of the LOFAR data indicates variations in CRRL
optical depth on angular scales significantly smaller than the beam sizes used in the observational data from the previous literature studies quoted here. Nevertheless,
this example illustrates the power of CRRL observations to measure the physical conditions in diffuse interstellar clouds.

\subsubsection{Galactic Center Regions}

As a second example, we analyze observations of clouds detected towards regions in the galactic plane \citep{erickson1995}. In view of the scarceness, low
spatial resolution and limited frequency coverage of the data available in the literature, our results should be taken with care and considered illustrative.
We chose two regions with good signal to noise measurements (SNR$>10$). In Table 1, we show the line parameters for $\mathrm{C}441\alpha$ and $\mathrm{C}555\beta$
lines from \citet{erickson1995} with a beam size of $4^\circ$.

\begin{table}[ht]
\caption{Selected values for $\mathrm{C}n\alpha$ and $\mathrm{C}n\beta$ lines for regions observed by \citet{erickson1995}.}
\begin{tabular}{ccccc}
\hline
\hline
Name &   $\tau(441\alpha)$ & $\Delta v(441\alpha)$ &   $\tau(555\beta)$ & $\Delta v(555\beta)$\\
     &     $\times10^{-3}$  & $\mathrm{km~s^{-1}}$&     $\times10^{-3}$  & $\mathrm{km~s^{-1}}$\\
G000.0+0  & 0.73$\pm$0.03 & 24$\pm$1 & 0.35$\pm$0.03 & 24$\pm$2  \\
G002.0-2  & 0.97$\pm$0.08 &  9$\pm$1 & 0.75$\pm$0.04 & 25$\pm$2  \\
\hline
\end{tabular}
\end{table}

\begin{figure}[ht]
\includegraphics[width=0.5\columnwidth]{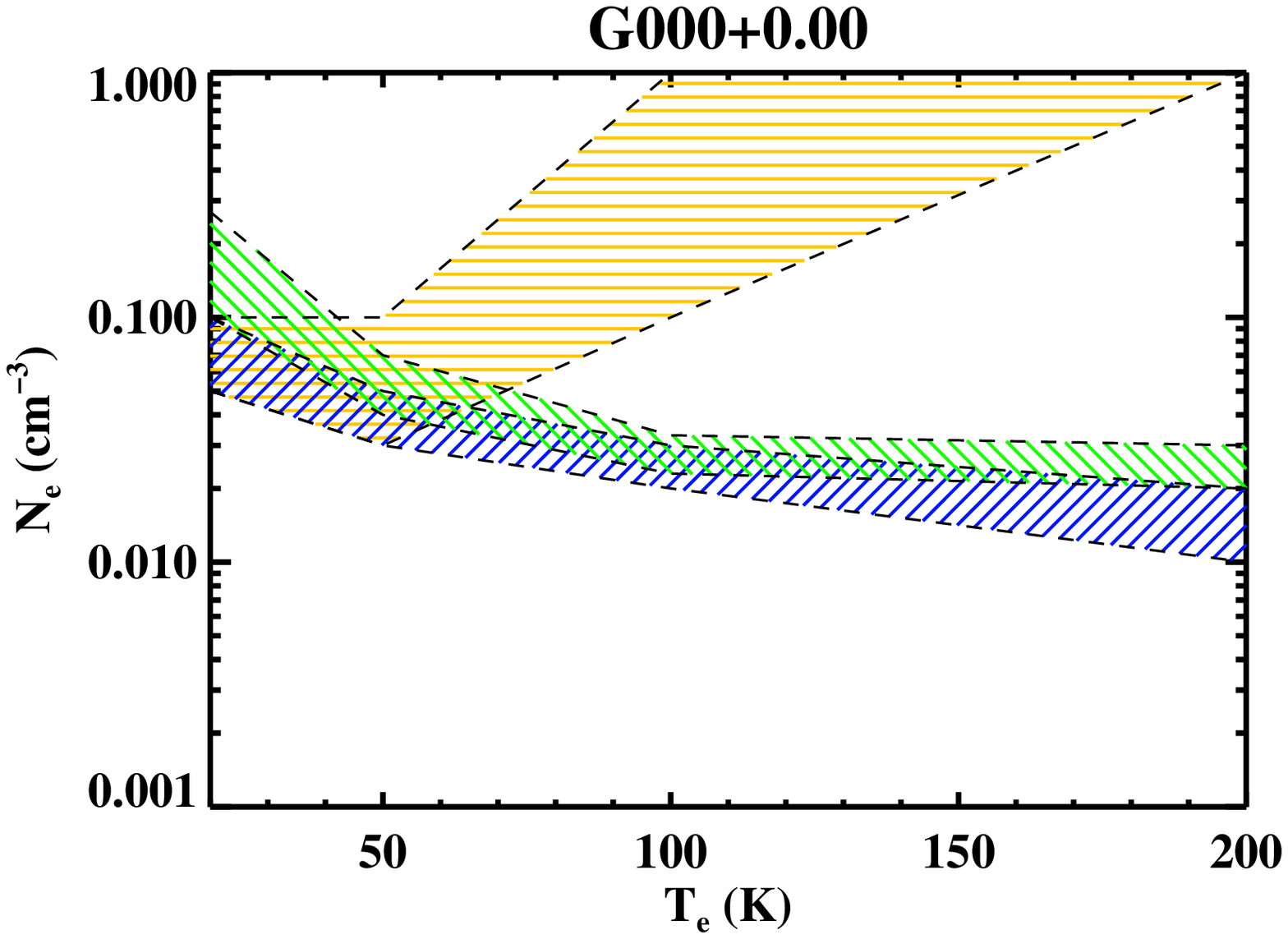}
\includegraphics[width=0.5\columnwidth]{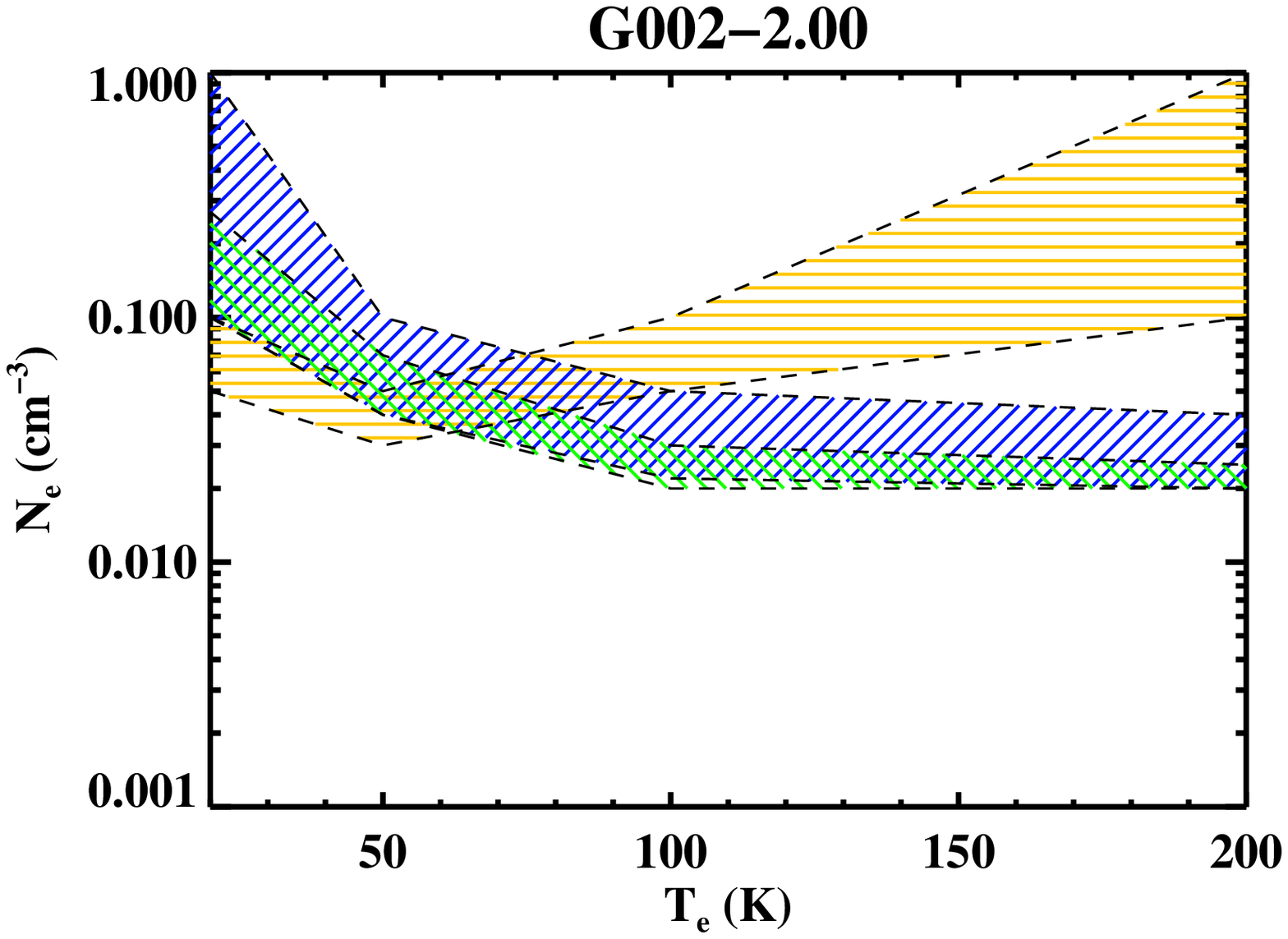}
\caption{Same as Figure~\ref{fig_contourcasa} for regions towards the Galactic center (data from \citealt{erickson1995}). The $\alpha$ to $\beta$
ratio constraints is shown as a blue region. The constraints derived from $n_t$ are shown as a green zone.  In addition, we have added the constraint
from CRRL to [CII] 158 $\micron$ ratio as the orange shaded zone.\label{fig_contourGC}}
\end{figure}

In Figure \ref{fig_contourGC}, we summarize the constraints imposed by the integrated $\alpha$ to $\beta$ line ratio as a blue zone,
the transition from emission to absorption ($n_t$ level) as a green zone (we estimate to be $350<n_t<400$) and the integrated
line-to-continuum to the $I(158~\mu m)$ ratio as the orange zone.

From the line widths towards the lines of sight in Table 1, an upper limit to the density can be estimated by assuming pure collisional broadening, as shown
in Section \ref{section_lineprofile}. The upper limits on density are $N_e\leq 1.5~\mathrm{cm^{-3}}$ for G000.0+0 and $N_e\leq 0.5~\mathrm{cm^{-3}}$ for G002.0-2. The constraint is
even more strict when considering that part of the broadening must be produced by the Galactic radiation field. Assuming no collisional broadening,
the upper limits on the background temperatures for the regions are ${T_0}\leq 4\times10^4~\mathrm{K}$ for G000.0+0 and ${T_0}\leq 1.5\times10^4~\mathrm{K}$ for G002.0-2.
These are strict upper limits as the observations from \citet{erickson1995} were performed with large beams and the observed lines are likely produced by
several ``clouds'' in the beam.

We estimate the value for $I(158~\mathrm{\mu m})$ to be $8-12\times10^{-5}~\mathrm{erg~s~cm^{-2}~sr^{-1}}$ from \emph{COBE} data \citep{bennett1994}.
Since the data from \citet{erickson1995} is for the $\mathrm{C}441\alpha$ line, we created a diagnostic plot similar to that in Figure \ref{fig_diagplotdensity}
for the level 441. We obtain for G000.0+0, a value for $T_e$ between $20~\mathrm{and}~60~\mathrm{K}$ and $N_e$ between $4\times10^{-2}~\mathrm{and}~1\times10^{-1}~\mathrm{cm^{-3}}$.
For G002.0-2, we obtain $T_e=20~\mathrm{to}~80~\mathrm{K}$ and $N_e=4\times10^{-2}-1\times10^{-1}~\mathrm{cm^{-3}}$. With these values and using
Equation \ref{eqn_integlinetocontx} we determine lengths of 2 to 19~pc for G000.0+0 and 1 to 9~pc for G002.0-2. Assuming that the electrons are
provided by carbon ionization and adopting a carbon gas phase abundance of $1.6\times 10^{-4}$, we derive thermal pressures between 5000 and 37500~$\mathrm{K~cm^{-3}}$
for G000.0+0 and between 5000 and 50000~$\mathrm{K~cm^{-3}}$ for G002.0-2. Strictly speaking the values from COBE include emission produced in the
neutral and warm components along the lines of sight. Since CRRLs are expected to be produced predominantly in cold clouds, the here determined ratio between
the CRRL and the [CII] line can be underestimated. However, \citet{pineda2013} showed that the contribution from ionized gas to the [CII] line is $\sim 4\%$ towards the inner Galaxy.
It is clear from Figure \ref{fig_contourGC} that the $n_t$ level (green zone in the plots) and the integrated $\alpha$-to-$\beta$ line ratio
provide similar constrains in the $N_e$ vs. $T_e$ plane. By far, the strongest constraint comes from $n_t$, since the errors in the measurements do not provide strong limits on the $\alpha$-to-$\beta$ line ratio. 
As the error bars are rather large, the derived constraints -- given above -- are not very precise. Nevertheless, the inherent power of CRRL for quantitative studies
of diffuse clouds in the ISM is quite apparent.

\subsection{Discussion}

As the examples of Cas~A and G000.0+0 and G002.0-2 show, a large amount of relevant physical information on the properties of the clouds
can be obtained from CRRL measurements, despite the scarceness of the data used here. The $\alpha$-to-$\beta$ line ratios can provide powerful
constraints as long as the frequency observed is higher than 30 MHz. As illustrated by our Cas~A example, the CRRL ratio plots
can be extremely useful in constraining the electron density and temperature, and lines with a large separation in terms
of quantum number are expected to be the most useful ratios. As illustrated in Figure \ref{fig_integtauratio},
ratios between levels around 300 and 500 can provide direct constraints or indirect constraints by using, in addition, the $n_t$ value.
An advantage of using ratios is that they only depend on the local conditions and beam filling factors are of little concern.

Although we consider our examples illustrative the determined values for $T_e$ and $n_e$ are within the values expected from theory
(e.g. \citealt{wolfire2003, kim2011}) and  HI 21 cm observations \citep{heilesandtroland2003b}. Moreover, the derived thermal pressures
agree well with those derived from CI UV lines in the local ISM \citep{jenkins1979,jenkins1983,jenkins2011}.

\section{Summary and Conclusions}\label{section_conclusions}

In this paper we have analyzed carbon radio recombination line observations. Anticipating the LOFAR CRRL survey, we focus our
study in the low frequency regime, corresponding to transitions between lines with high principal quantum number. We have studied
the radiative transfer of recombination lines and the line broadening mechanisms in the most general form.

Our results show that line widths provide constraints on the physical properties of the gas. At high frequencies the observed line
widths provide limits on the gas temperature and on the turbulent velocity of the cloud. At low frequencies, observed line widths
provide constraints on the electron density of the intervening cloud and on the radiation field that the cloud is embedded in.
Using the departure coefficients obtained in Paper~I, we analyzed the behavior of the lines under the physical conditions of the diffuse ISM. 
Integrated optical depths provide constraints on the electron density, electron temperature and the emission measure or size of the cloud. The
use of CRRLs together with [CII] at 158~$\mu\mathrm{m}$ can constrain the temperature.

As an illustration of the use of our models, we have analyzed existing data in low frequency CRRLs towards Cas~A and the inner galaxy to derive
physical parameters of the absorbing/emitting clouds \citep{payne1994,stepkin2007,erickson1995}.

Our models predict that detailed studies of CRRLs should be possible with currently available instrumentation. By using realistic
estimates for the properties of the diffuse ISM we obtain optical depths that are within the capabilities of LOFAR and of the future Square
Kilometer Array \citep{oonk2015a}. Given the clumpy nature of the ISM, we encourage observations with high angular resolution. Observations
with large beams are biased towards line of sights with large optical depth and narrow lines, and these happen to be clouds of low density
for a given temperature. High spectral resolution is also encouraged in order to distinguish multiple components along the line of sights.
Once the temperature and the density have been determined, the observed intensities yield the C$+$ column density which
can be combined with the HI column density from 21 cm observations to determine the gas phase carbon abundance.

The main conclusions of our work are:

1) CRRLs provide a powerful probe of the physical conditions of diffuse interstellar clouds.

2) Meaningful constraints on gas properties can be derived from combining information on the location of the transition
from emission to absorption, $\alpha$-to-$\beta$ ratios and $\alpha$-line ratios spread in frequency.
Further limits are provided by the low frequency line width.

3) Comparison of CRRLs with [CII] 158 $\mu \mathrm{m}$ line measured by COBE \citep{bennett1994}, BICE \citep{nakagawa1998}
and Herschel (GOT C+;\citealt{pineda2013}); in addition to new observations with the German Receiver for Astronomy at Terahertz Frequencies
(GREAT; \citealt{heyminck2012}) on board of SOFIA, will provide important constraints primarily on the temperature, but also aid in further
constraining the density and size of diffuse clouds.

\clearpage
\appendix

\section{List of Symbols}\label{appendix_listofsymbols}
 
\begin{center}
\begin{longtable}{lp{100mm}}
\caption{List of Symbols\label{table_listofsymbols}}\\ 
\multicolumn{2}{l}{}\\
\hline
\hline
Symbol & Descritpion \\
\hline
\endfirsthead
\multicolumn{2}{l}
{\tablename\ \thetable\ -- \textit{Continued from previous page}} \\
\hline
\hline
Symbol &  \\
\hline
\endhead
\hline \multicolumn{2}{r}{\textit{Continued on next page}} \\
\endfoot
\endlastfoot
 $A_{n'n}$ & Einstein coefficient for spontaneous transitions\\
 $A_{3/2,1/2}$& Spontaneous transition rate of the carbon fine structure line ${^2}P_{3/2}$-${^2}P_{1/2}$\\
 $a(T_e)$& Fitting coefficient for collisional broadening\\
 $B_{nn'}$ & Einstein coefficient for stimulated transition\\
 $b_n$& Departure coefficient for level n\\
 $B_{\nu}(T)$ & Planck function at frequency $\nu$ for a temperature $T$\\
 $C_{n'n}$& Rates for energy changing collisions between level $n'$ and $n$\\
 $\mathrm{C}n\alpha$& Carbon recombination line with $\Delta n=1$\\
 $c$& Speed of light\\ 
 $EM_{\mathrm{C+}}$& Emission measure of carbon ions\\
 $h$& Planck constant\\
 $I_0(\nu)$& Intensity of the background continuum\\
 $I_\nu^{line}$& Intensity of the line\\
 $I_\nu^{cont}$& Intensity of the continuum\\
 $I_{158}$& Intensity of the fine structure line of carbon at 158~$\mathrm{\mu m}$\\
 $j_\nu^l$ & line emission coefficient\\
 $j_\nu^c$ & continuum emission coefficient\\
 $k$& Boltzmann constant\\
 $k_\nu^l$ & line absorption coefficient\\
 $k_\nu^c$ & continuum absorption coefficient\\
 $L$& Pathlength of the cloud\\
 $M(\Delta n)$& Approximation factor for the oscillator strength, as given by \citet{menzel1968}\\
 $m_C$& Mass of a carbon atom\\
 $N_{3/2}^+$& Level population of carbon ions in the ${^2}P_{3/2}$ core\\
 $N_{\mathrm{C}+}$& Number density of carbon ions\\
 $N_e$& Electron density\\
 $\mathcal{N}_\mathrm{C^+}$& Carbon column density\\
 $\mathcal{N}_\mathrm{H}$& Hydrogen column density\\
 $n$& Lower principal quantum number\\
 $n'$& Upper principal quantum number\\
 $n_t$& Level where the observed lines transition from emission to absorption\\
 $R$& Ratio between the fine structure (${^2}P_{3/2}$-${^2}P_{1/2}$) level population and the fine structure level population in LTE\\
 $Ry$& Rydberg constant\\
 $T_0$& Temperature of power law background spectrum at frequency $\nu_0$\\
 $t_n$& Ratio of radiation to collisional broadening\\
 $T_X$ & Excitation temperature\\
 $T_e$& Electron temperature\\
 $\langle v_{RMS} \rangle$& RMS turbulent velocity\\
 $\alpha_{1/2}$& Fraction of carbon ions in the ${^2}P_{1/2}$ level\\
 $\alpha_{pl}$& Exponent of the power law background spectrum\\
 $\beta_{n n'}$& Correction factor for stimulated emission\\
 $\beta_{158}$& Correction for simulated emission to the [CII] fine structure line ${^2}P_{3/2}$-${^2}P_{1/2}$\\
 $\gamma_c(T_e)$& Fitting coefficient for collisional broadening\\
 $\gamma_e$& De-excitation rate for carbon ions in the ${^2}P_{3/2}$ core due to collisions with electrons\\ 
 $\gamma_H$& De-excitation rate for carbon ions in the ${^2}P_{3/2}$ core due to collisions with hydrogen atoms\\
 $\Delta n$& $n'-n$, difference between the upper and lower principal quantum number\\
 $\Delta \nu_D$& Doppler width\\
 $\Delta \nu_{rad}$& Radiation broadening\\
 $\Delta \nu_{col}$& Collisional broadening\\
 $\nu$& Frequency of a transition\\
 $\eta$& Correction factor to the Planck function due to non-LTE level population\\
 $\tau_{158}$& Optical depth for the [CII] fine structure line ${^2}P_{3/2}$-${^2}P_{1/2}$\\
 $\tau_\nu^l$& Optical depth of the line\\
 $\tau_\nu^c$& Optical depth of the continuum\\
 $\tau_\nu^{total}$& Sum of $\tau_\nu^l$ and $\tau_\nu^c$\\
 $\phi(\nu)$& Line profile\\
 $\nu_0$& Reference frequency for the power law background spectrum\\
 ${\chi_n}$ & ${hc Z^2 Ry /n^2 k T_e}$ \\
\hline
\end{longtable} 
\end{center}

\section{Radiative Transfer}\label{appendix_radtransfer}
The radiative transfer equation for a line in the plane parallel approximation is given by:
\begin{eqnarray}
\frac{dI_\nu}{dx}(x)&=& -k_\nu(x) I_\nu(x)+j_\nu(x)\\
k_\nu&=&k_\nu^l+k_\nu^c\\
j_\nu&=&j_\nu^l+j_\nu^c,
\end{eqnarray}
\noindent where $k_\nu^l$ is the line absorption coefficient, $k_\nu^c$ is the continuum absorption coefficient, $j_\nu^l$ is the line emission coefficient,
$j_\nu^c$ is the continuum emission coefficient and $I_\nu(x)$ is the specific intensity of a nebula at a frequency $\nu$ as a function of depth in the cloud $x$.
The line absorption and emission coefficients are given by:
\begin{eqnarray}\label{eqn_emmcoeff}
j_\nu^l &=& \frac{h \nu}{4 \pi} A_{n'n} N_{n'} \phi(\nu),\\
k_\nu^l &=& \frac{h \nu}{4 \pi} \left( N_{n} B_{nn'}-N_{n'} B_{n'n}  \right) \phi(\nu),
\end{eqnarray}
\noindent where $N_{n'}$ is the level population of a given upper level and $N_{n}$ is the level population of the lower level;
$\nu$ is the frequency of the transition and $A_{n'n}$, $B_{n'n}(B_{nn'})$ are the Einstein coefficients for spontaneous and stimulated
emission (absorption), related to each other by:
\begin{eqnarray}
A_{n'n}&=& \frac{2 h\nu^3}{c^2} B_{n'n},\\
B_{nn'}&=& \frac{\omega_{n'}}{\omega_n} B_{n'n}.
\end{eqnarray}
The factor $\phi(\nu)$ in Equation~\ref{eqn_emmcoeff} is the normalized line profile ($\int \phi(\nu) \mathrm{d}\nu=1$). The effects on the emission
are analyzed in Section \ref{section_lineprofile}. Here, we assume that $j_\nu$ is evaluated at the line center where the frequency of the transition is $\nu_0$
and omit the $\phi(\nu_0)$ factor. Note that due to the normalization, $\phi(\nu_0)<1$.
Under thermodynamic equilibrium the level population of a level $n$ ($N_n(LTE)$) is given by:
\begin{eqnarray}
N_{n}(LTE)&=&N_e N_{ion}\left(\frac{h^2}{2 \pi m_e k T_e}\right)^{1.5} \frac{\omega_{n}}{2\omega(i)} e^{\chi_n}, \chi_n=\frac{hc Z^2 Ry}{n^2kT_e},
\end{eqnarray}
\noindent where $N_e$~is the electron density in the nebula, $T_e$ is the electron temperature, $N_{ion}$~is the ion density, $m_e$~is the electron mass,
$h$ is the Planck constant, $k$ is the Boltzmann constant, $c$ is the speed of light, $Ry$ is the Rydberg constant and $\omega_{n}$~is the statistical weight
of the level $n$~($\omega_{n}=2n^2$, for hydrogen).
In the ISM, levels can be out of local thermodynamic equilibrium (Paper~I). The level population can then be described by the departure coefficients
$b_n = N_n/N_n(LTE)$, i. e. the ratio of the level population of a given level to its LTE value.
From the definitions of $j_\nu^l$ and $k_\nu^l$, we can write the emission and absorption coefficients in terms of the departure coefficients:
\begin{eqnarray}
j_\nu^l &=& j_\nu^l(LTE) b_n,\\
k_\nu^l&=& \frac{h \nu}{4 \pi} \left( b_n N_{n}(LTE) B_{nn'}- b_{n'}N_{n'}(LTE)B_{n'n}  \right),\\
&=& k_\nu^l(LTE) b_n \frac{1 - \frac{b_{n'}}{b_n} e^{-h\nu/kT_e}}{1 - e^{-h\nu/kT_e}}.
\end{eqnarray}
\noindent The correction factor for stimulated emission/absorption, $\beta_{nn'}$, is:
\begin{eqnarray}
\beta_{nn'}&=& \frac{1-\frac{b_{n'}}{b_n} e^{-h\nu/kT_e}}{1-e^{-h\nu/kT_e}}.
\end{eqnarray}
Deviations from equilibrium can be also described in terms of the excitation temperature ($T_X$) of a transition, defined as:
\begin{eqnarray}
\frac{N_{n'}/\omega_{n'}}{N_{n}/\omega_{n}}&=& \exp\left(\frac{-h \nu}{kT_X}\right).
\end{eqnarray}
\noindent It is easy to see that $T_X$ is related to $\beta_{nn'}$ by:
\begin{eqnarray}
\beta_{nn'}&=& \frac{1-e^{-h\nu/kT_X}}{1-e^{-h\nu/kT_e}}.
\end{eqnarray}
\noindent Clearly, under LTE conditions the excitation temperature approaches the value of the electron temperature, i.e. $T_X=T_e$. The description of the level
population in terms of $T_X$ is useful to explain the behavior of the lines as we show in Section \ref{section_results}.
For a homogeneous cloud, the radiative transfer equation can be solved. At a given frequency, the observed flux has contributions from both the line and
the continuum; which can be written as:
\begin{eqnarray}
I_\nu^{total} &=& \frac{j_\nu^c+j_\nu^l}{k_\nu^c+k_\nu^l} \left[1-e^{-(\tau_\nu^c+\tau_\nu^l)}\right]+I_0(\nu) e^{-(\tau_\nu^c+\tau_\nu^l)},\\
I_\nu^{c} &=& \frac{j_\nu^c}{k_\nu^c} (1-e^{-\tau_\nu^c})+I_0(\nu) e^{-\tau_\nu^c},
\end{eqnarray}
\noindent where a background continuum source, $I_0(\nu)$ has been introduced. The coefficients $\tau_\nu^x=\int k^x_\nu(s) ds$
are the optical depth for $x$ of either the continuum or the line. Assuming homogeneity $\tau_\nu^x= k^x_\nu L$, where $L$ is the length along the line of sight
of the cloud, we can separate the contribution from the line itself since it is given by:
\begin{eqnarray}
I_\nu^{line}&=& I_\nu^{total}- I_\nu^{c}\\
I_\nu^{total}- I_\nu^{continuum} &=& \frac{j_\nu^c+j_\nu^l}{k_\nu^c+k_\nu^l} (1-e^{-\tau_\nu^{total}})+I_0(\nu) e^{-\tau_\nu^{total}}\\
&&- \frac{j_\nu^c}{k_\nu^c}(1-e^{-\tau_\nu^c})-I_0(\nu) e^{-\tau_\nu^c} \nonumber\\
\tau_\nu^{total} &=&\tau_\nu^c+\tau_\nu^l.
\end{eqnarray}
We can write the line contribution in terms of the source function ($S_\nu$) by using Kirchoff's law [$j_\nu=\kappa_\nu B_\nu(T_e)$, with $B_\nu(T_e)$
the Planck function]:
\begin{eqnarray}
S_\nu &=& \frac{j_\nu^c+j_\nu^l}{k_\nu^c+k_\nu^l} \nonumber\\
      &=& \left[\frac{k_\nu^c+b_{n'} k_\nu^l(LTE)}{k_\nu^c+b_n \beta_{nn'}k_\nu^l(LTE)}\right]B_\nu(T_e).
\end{eqnarray}

We identify a correction factor to the Planck function for departures from LTE:
\begin{eqnarray}
\eta&=&\frac{k_\nu^c+b_{n'} k_\nu^l(LTE)}{k_\nu^c+b_n \beta_{nn'}k_\nu^l(LTE)},
\end{eqnarray}
\noindent as in e.g. \citet{strelnitski1996} and \citet{gordon2009}.

With these definitions, we can write:
\begin{eqnarray}
I_\nu^{line}&=& \eta B_\nu(T_e)(1-e^{-\tau_\nu^{total}}) -B_\nu(T_e)(1-e^{-\tau_\nu^c})+ \nonumber\\
&&+I_0(\nu)e^{-\tau_\nu^c}\left(e^{-\tau_\nu^l}-1\right),
\end{eqnarray}
\noindent and the intensity of a line relative to the continuum is:
\begin{eqnarray}\label{eq_linecontgeneral_app}
\frac{I_\nu^{line}}{I_\nu^{cont}} &=& \frac{\eta B_\nu(T_e) (1-e^{-\tau_\nu^{total}})+I_0(\nu) e^{-\tau_\nu^{total}}}{B_\nu(T_e)(1-e^{-\tau_\nu^c}) + I_0(\nu) e^{-\tau_\nu^c}} -1,
\end{eqnarray}
In the absence of a background radiation field ($I_0=0$) this reduces to:
\begin{eqnarray}
\frac{I_\nu^{line}}{I_\nu^{cont}}&=& \frac{\eta  (1-e^{-\tau_\nu^{total}})}{(1-e^{-\tau_\nu^c})}-1.
\end{eqnarray}

In Section \ref{section_lineprofile}, we showed that under the conditions of the diffuse ISM the line profile is expected to be Lorentzian in shape,
and, at the line center, $\phi(\nu_0)=2/\pi\Delta\nu_L$ with $\Delta \nu_L$ the full width at half maximum (FWHM) of the line. This sets a range
of physical parameters for which the approximation $|\tau_\nu^l| \ll1$ is valid.

\subsection{Doppler and Lorentzian broadening}\label{appendix_dopbroadening}
Doppler broadening occurs due to turbulent motions in the gas and  thermal motions is given by a Gaussian
distribution with a Doppler width \citep{rybicki1986}:
\begin{eqnarray}
\Delta \nu_D= \frac{\nu_0}{c}\sqrt{\frac{2kT}{m_{atom}}+\langle v_{RMS} \rangle^2},
\end{eqnarray}
\noindent where $m_{atom}$ is the mass of the atom and $\langle v_{RMS} \rangle$ is the RMS turbulent velocity.
The line profile as a function of frequency is given by the expression:
\begin{eqnarray}
\phi_\nu^G(\nu)=\frac{1}{\Delta \nu_D \sqrt{\pi}} \exp{-\left(\frac{\nu-\nu_0}{\Delta \nu_D} \right)^2}.
\end{eqnarray}
\noindent With this definition the FWHM is $\Delta \nu_D(\mathrm{FWHM})= 2\sqrt{\mathrm{ln(2)}}\Delta \nu_D$.
Note that Doppler broadening is dominated by turbulence for $T_e~<~60.5~(m_{atom}/m_H) (\langle v_{RMS}\rangle/\mathrm{km~s^{-2}})^2~\mathrm{K}$,
here $m_H$ is the mass of a proton.

The Lorentzian width of a line produced by a transition from a level $n'$ to $n$ is related to the net transition out of the levels \citep{shaver1975,rybicki1986}:
\begin{eqnarray}
\Gamma_{n'n}&=& \Gamma_{n'}+\Gamma_{n},\\
\Gamma_{n'}&=& \sum_{n<n'} A_{n'n}+\sum_{n\neq n'} N_e C_{n'n}+\sum_{n \neq n'} B_{n'n} I_\nu,\\
&=& \Gamma_{natural}+\Gamma_{collisions}+\Gamma_{radiation}
\end{eqnarray}
\noindent with an analogous formula for $\Gamma_{n}$. Here, we have to consider
collisions with electrons and transitions induced by an external radiation field. This produces a Lorentzian line profile:
\begin{eqnarray}
\phi_\nu^L(\nu)=\frac{\gamma}{\pi} \frac{1}{(\nu-\nu_0)^2+\gamma^2},
\end{eqnarray}
\noindent the FWHM is $\Delta \nu_L(\mathrm{FWHM})= 2 \gamma$. The width $\gamma$ of a line transition between levels $n$ and $n'$ is given by
$\gamma=(\Gamma_n+\Gamma_{n'})/4\pi$. For transitions between lines with $n\approx n'$ we have $\Gamma_n\approx\Gamma_{n'}$ and $\gamma\approx\Gamma_{n}/2\pi$.

In the most general case, the line profile is given by the Voigt profile, i.e. the convolution of the Gaussian and the Lorentzian profile, e. g. \citet{gordon2009}:
\begin{eqnarray}
\phi_\nu^V(\nu)=\int_{-\infty}^{\infty} \phi_\nu^L(\nu) \phi_\nu^G(\nu) \mathrm{d}\nu.
\end{eqnarray}
\noindent This can be written in terms of the Voigt function [$H(a,u)$] by using the proper normalization:
\begin{eqnarray}
\phi_\nu^V( \nu)&=&\frac{1}{\Delta \nu_D \sqrt{\pi}} H(a,u) \\
H(a,u)&=&\frac{a}{\pi} \int_{-\infty}^{\infty} \frac{e^{-y^2} \mathrm{d}y}{a^2+(y-u)^2},
\end{eqnarray}
\noindent with $a=\gamma/\Delta \nu_D$ and $u=(\nu-\nu_0)/\Delta \nu_D$.
The FWHM of the Voigt profile can be approximated by:
\begin{eqnarray}
\Delta \nu_V(\mathrm{FWHM})=0.5346 \Delta \nu_L(\mathrm{FWHM})+\sqrt{0.2166 \Delta \nu_L(\mathrm{FWHM})^2+\Delta \nu_D(\mathrm{FWHM})^2}.
\end{eqnarray}

\subsection{Collisional/Stark broadening}\label{appendix_colbroadening}

Collisions with electrons produce line broadening:
\begin{eqnarray}
\Delta \nu_{col} = \frac{2}{\pi}\sum_{n\neq n'} N_e C_{n'n},
\end{eqnarray}
\noindent where $C_{n'n}$ is the collision rate for electrons induced transitions from level $n'$ to $n$, and
$N_e$ is the electron density. For levels $n > 100$ we fitted the following function to depopulating collisions:
\begin{eqnarray}\label{eqn_fitcolapp}
\sum_{n\neq n'} N_e C_{n'n} =  N_e 10^a \times n^{\gamma_c}.
\end{eqnarray}
\noindent Values for $a$ and $\gamma_c$ are given in Table~\ref{table_col}.
The values used here agree with those from Shaver at low temperatures, but at temperatures larger than about $1000~\mathrm{K}$ they can differ
by factors larger than about $4$. The values presented here agree well with those of \citet{griem1967} at large temperatures.
At low frequencies, collisional broadening is large and dominates over Doppler broadening in the absence of a background radiation field.
As can be seen from the dependence on the electron density, clouds with higher densities have broader lines than those with lower
densities at a given level $n$.

\begin{center}
\begin{table}
\caption{Coefficients for Equation~\ref{eqn_fitcol}.}\label{table_col}
\begin{tabular}{ccc}
\hline
\hline
$T_e~\mathrm{(K)}$ & $a$ & $\gamma_c$ \\
\hline
10  & -10.97  &  5.482 \\
20  & -10.67  &  5.435 \\
30  & -10.49  &  5.407 \\
40  & -10.37  &  5.386 \\
50  & -10.27  &  5.369 \\
60  & -10.19  &  5.354 \\
70  & -10.12  &  5.341 \\
80  & -10.06  &  5.329 \\
90  & -10.01  &  5.318 \\
100 & -9.961  &  5.308 \\
200 & -9.620  &  5.228 \\
300 & -9.400  &  5.170 \\
400 & -9.234  &  5.122 \\
500 & -9.085  &  5.077 \\
600 & -8.969  &  5.041 \\
700 & -8.869  &  5.009 \\
800 & -8.780  &  4.980 \\
900 & -8.701  &  4.953 \\
1000& -8.630  &  4.929 \\
2000& -8.272  &  4.806 \\
3000& -8.009  &  4.706 \\
4000& -7.834  &  4.636 \\
5000& -7.708  &  4.583 \\
6000& -7.613  &  4.542 \\
7000& -7.538  &  4.509 \\
8000& -7.477  &  4.482 \\
9000& -7.427  &  4.458 \\
10000& -7.386 &  4.439 \\
20000& -7.181 &  4.329 \\
30000& -7.113 &  4.281 \\
\hline
\end{tabular}
\end{table}
\end{center}

\section{Radiation broadening}\label{appendix_radbroadening}
The depopulation of a given level $n'$ due to stimulated transitions is given by:
\begin{eqnarray}
\Gamma_{n'}^{radiation} = \sum_{n \neq n'} B_{n'n} I_\nu.
\end{eqnarray}
\noindent where $B_{n'n}$ is the Einstein $B$ coefficient for stimulated transitions from level $n'$ to $n$, and
$I_\nu$ is an external radiation field. 

We can write the Einstein $B_{n'n}$ coefficients in terms of the Einstein $A_{n'n}$ coefficients:
\begin{eqnarray}
B_{n+\Delta n,n} I_\nu=\frac{c^2}{2 h \nu^3} A_{n+\Delta n,n} I_\nu,
\end{eqnarray}
\noindent (e.g. \citealt{shaver1975} and \citealt{gordon2009}) where we have used the $n'=n+\Delta n$.
Assuming a power-law like radiation field, with temperature $T_R=T_0 (\nu/\nu_0)^{\alpha_{pl}}$ we can write:
\begin{eqnarray}\label{eqn_bnnann}
B_{n+\Delta n,n} I_\nu=\frac{ k T_0}{h \nu_0^{\alpha_{pl}}} A_{n+\Delta n,n} \nu^{\alpha_{pl}-1}.
\end{eqnarray}
The Einstein $A$ coefficient can be written in terms of the oscillator strength, $f_{n,n+\Delta n}$ (e.g. \citealt{shaver1975}):
\begin{eqnarray}
A_{n+\Delta n,n} = \frac{8 \pi^2 e^2 \nu^2}{m_e c^3} \left(\frac{n}{n+\Delta n} \right)^2 f_{n, n+\Delta n},
\end{eqnarray}
Replacing in Equation~\ref{eqn_bnnann} leads to:
\begin{eqnarray}\label{eqn_appbnnp}
B_{n+\Delta n,n} I_\nu&=& \frac{k T_0}{h \nu_0^{\alpha_{pl}}} \frac{8 \pi^2 e^2 \nu^2}{m_e c^3} \left(\frac{n}{n+\Delta n} \right)^2 f_{n, n+\Delta n} \nu^{\alpha_{pl}-1}, \nonumber\\
&=& \frac{8 \pi^2 e^2}{ m_e c^3} \left( \frac{k T_0}{h\nu_0^{\alpha_{pl}}}\right)  \left(\frac{n}{n+\Delta n} \right)^2 f_{n, n+\Delta n} \nu^{\alpha_{pl}+1}.
\end{eqnarray}

\citet{menzel1968} gives a simple approximation for computing the oscillator strength:
\begin{eqnarray}
\frac{f_{n+\Delta n,n}}{n} \approx M(\Delta n) \left(1+ \frac{3}{2} \frac{\Delta n}{n} \right),
\end{eqnarray}
\noindent with $M(\Delta n)=4/3 J_{\Delta n}(\Delta n) J'_{\Delta n}(\Delta n)/\Delta n^2$, where $J_{\Delta n}(\Delta n)$ is
the Bessel function of order $\Delta n$. The $M(\Delta n)$ can be approximated by  $M(\Delta n)\approx 0.1908/\Delta n^3$ to
less than 16\% in accuracy for $\Delta n=5$, and to better than 3\% accuracy, by changing the exponent from 3 to 2.9. 
The values for $M(\Delta n)=0.1908,~0.02633,~0.008106,~0.003492,~0.001812$~for $\Delta n=1,~2,~3,~4,~5$.

The frequency of a line in the hydrogenic approximation is given by:
\begin{eqnarray}
\nu_{n+\Delta n,n} &=&Ry c Z^2\left( \frac{1}{n^2}-\frac{1}{\left(n+\Delta n \right)^2}\right),\\
&\approx& 2 Ry c Z^2 \frac{ \Delta n}{n^3}\left(1- \frac{3}{2} \frac{\Delta n}{n} \right),
\end{eqnarray}
(e. g. \citealt{shaver1975,gordon2009}). Replacing $\nu$ in Equation~\ref{eqn_appbnnp} and for $Z=1$, we obtain:
\begin{eqnarray}
B_{n+\Delta n,n} I_\nu&=& \frac{8 \pi^2 e^2}{ m_e c^3} \left( \frac{k T_0}{h \nu_0^{\alpha_{pl}}}\right)\left(\frac{n}{n+\Delta n} \right)^2 M(\Delta n) \left(1+ \frac{3}{2} \frac{\Delta n}{n} \right)n \left[2 Ry c \frac{ \Delta n}{n^3}\left(1- \frac{3}{2} \frac{\Delta n}{n} \right)\right]^{\alpha_{pl}+1}, \nonumber \\
&\approx& \frac{8 \pi^2 e^2 (2 Ry c)^{\alpha_{pl}+1} k T_0}{m_e c^3 h \nu_0^{\alpha_{pl}}} M(\Delta n) n \left(\frac{\Delta n}{n^3}\right)^{\alpha_{pl}+1}, \nonumber
\end{eqnarray}
\noindent for $\Delta n /n \ll 1$. Rearranging the expression we arrive to:
\begin{eqnarray}\label{eqn_appc}
B_{n+\Delta n,n} I_\nu&=& \frac{8 \pi^2 e^2 (2 Ry c )^{\alpha_{pl}+1} k T_0}{m_e c^3 h \nu_0^{\alpha_{pl}}} M(\Delta n)n^{-3\alpha_{pl}-2} \Delta n^{\alpha_{pl}+1},\nonumber \\
&=& \frac{8 \pi^2 e^2 (2 Ry c )^{\alpha_{pl}+1} k T_0}{m_e c^3 h \nu_0^{\alpha_{pl}}} 0.1908 n^{-3\alpha_{pl}-2} \Delta n^{\alpha_{pl}-2},\nonumber \\
&=& 2.137 \times 10^4 \left(\frac{6.578 \times 10^{15}}{\nu_0}\right)^{\alpha_{pl}+1} k T_0 \nu_0 n^{-3\alpha_{pl}-2} \Delta n^{\alpha_{pl}-2}.
\end{eqnarray}
\noindent Evaluating Equation~\ref{eqn_appc} for $T_0=22.6 \times 10^3~\mathrm{K},~\nu_0=30~\mathrm{MHz},~\alpha_{pl}=-2.55$ at $n=100$, and $\Delta n=1$,
we recover formula 2.177 of \citet{gordon2009}. Assuming $\alpha_{pl}=-2.6$ at a reference frequency of 100 MHz, Equation \ref{eqn_appc} is:
\begin{eqnarray}
B_{n+\Delta n,n} I_\nu&=&0.662 kT_0 n^{5.8} \Delta n^{-4.6} (\mathrm{s^{-1}}).
\end{eqnarray}
\noindent Other relevant values are for an optically thick thermal source $\alpha_{pl}=0$ and an optically thin thermal source
$\alpha_{pl}=-2.1$.

The broadening due to a radiation field in terms of the FWHM is:
\begin{eqnarray}
\Delta \nu_{rad}(FWHM)&=&\frac{2}{\pi} \sum_{\Delta n} B_{n+\Delta n,n} I_\nu,\\
&\approx& 5.819\times10^{-17} T_0 n^{5.8}(1+2^{-4.6}+3^{-4.6}),\\
&=& 6.096\times10^{-17} T_0 n^{5.8}~(\mathrm{s^{-1}}).
\end{eqnarray}

\subsection{The FIR fine structure line of $\mathrm{C+}$}\label{appendix_firline}
The beam averaged optical depth of the fine structure line of carbon ions for the transition ${^2}P_{1/2}-{^2}P_{3/2}$
is given by \citet{crawford1985,sorochenko2000}:
\begin{eqnarray}
\tau_{158} &=& \frac{c^2}{8 \pi \nu^2} \frac{A_{3/2,1/2}}{1.06 \Delta \nu} 2 \alpha_{1/2} \beta_{158} N_{\mathrm{C^+}} L,
\end{eqnarray} 
\noindent where $A_{3/2,1/2}=2.4\times10^{-6}~\mathrm{s^{-1}}$, $\nu$ is the frequency of the ${^2}P_{1/2}-{^2}P_{3/2}$ transition and
$\Delta \nu$ is the FWHM of the line. The $\alpha_{1/2}$ and $\beta_{158}$ defined by \citet{sorochenko2000} are:
\begin{eqnarray}
\alpha_{1/2} &=& \frac{1}{1+2 \exp(-92/T_e) R},\\
\beta_{158} &=& 1-\exp(-92/T_e)R.
\end{eqnarray}
\noindent The definition of $R$ is (\citealt{ponomarev1992,payne1994}; see also Paper~I):
\begin{eqnarray}\label{eq_payner_app}
R = \frac{N_e\gamma_e+N_H\gamma_H}{N_e\gamma_e+N_H\gamma_H+A_{3/2,1/2}},
\end{eqnarray}
\noindent where $\gamma_e~\mathrm{and}~\gamma_H$ are the de-excitation rates due to electrons and hydrogen atoms, respectively. For consistency
we used the same values as in Paper~I and neglected collisions with $\mathrm{H_2}$.

For the physical conditions considered here, we note that the FIR [CII] line is optically thin for hydrogen column densities of
$\sim1.2\times10^{21}~\mathrm{cm^{-2}}$. This corresponds to hydrogen densities of about $400~\mathrm{cm^{-3}}$ and electron densities
$6\times10^{-2}~\mathrm{cm^{-3}}$, assuming a length of the cloud of 1 pc and width of $2~\mathrm{km~s^{-1}}$.
The intensity of the [CII] 158 $\mu\mathrm{m}$~line in the optically thin limit is given by:
\begin{eqnarray}
I_{158}&=& \frac{h\nu}{4 \pi} A_{3/2,1/2} N^+_{3/2}\times L, \nonumber\\
       &=& \frac{h\nu}{4 \pi} \frac{A_{3/2,1/2} 2\exp(-92/T_e) R}{1+2 \exp(-92/T_e) R}\mathcal{N}_{\mathrm{C^+}},
\end{eqnarray}
\noindent with $N^+_{3/2}$ the number density of carbon ions in the $3/2$ state, $L$ the pathlength through the cloud along
the line of sight and $\mathcal{N}_{\mathrm{C^+}}$ the column density of carbon ions.
 Considering radiative transfer effects, the intensity of the line is given by:
\begin{eqnarray}
I_{158}&=& \frac{2h\nu_0}{\lambda^2} \frac{1.06 \Delta \nu(FWHM)}{e^{92/T_{158}}-1},
\end{eqnarray}
\noindent with $T_{158}$ defined as:
\begin{eqnarray} 
T_{158}=\frac{92}{\mathrm{ln}\left[(e^{92/T_e} e^{\tau_{158}}/R-1)/(e^{\tau_{158}}-1) \right]}.
\end{eqnarray}

\end{document}